\documentclass[preprint,showpacs,preprintnumbers,amsmath,amssymb,nofootinbib]{revtex4}
\usepackage{graphicx}
\usepackage{amsmath,amssymb}
\usepackage{bm}
\usepackage{color} 

\usepackage{bm}
\usepackage{ulem}
\def\lesssim{\ \raise.3ex\hbox{$<$}\kern-0.8em\lower.7ex\hbox{$\sim$}\ }
\def\gesim{\ \raise.3ex\hbox{$>$}\kern-0.8em\lower.7ex\hbox{$\sim$}\ }

\def\up{\uparrow}
\def\down{\downarrow}

\def\ket#1{|#1\rangle}

\def\br{{\bf r}}
\def\rB{{\rm B}}
\def\rI{{\rm I}}
\def\rBB{{\rm BB}}
\def\rIB{{\rm IB}}
\def\rex{{\rm ex}}
\def\intr{\int\!\!d^d\br}

\def\beq{ \begin{eqnarray} }
\def\eeq{ \end{eqnarray} }
\def\lk{\left(}
\def\rk{\right)}
\def\ltk{\left\{}
\def\rtk{\right\}}
\def\ldk{\left[}
\def\rdk{\right]}
\def\em#1{\it #1}

\begin{document}
\title{Polaron Problems in Ultracold Atoms: Role of {}{a Fermi Sea across Different Spatial Dimensions and Quantum Fluctuations of a Bose Medium}}

\author{Hiroyuki Tajima$^{1}$, Junichi Takahashi$^{2}$, Simeon I. Mistakidis$^{3}$, Eiji Nakano $^{1}$, and Kei Iida $^{1}$}
\affiliation{$^{1}$ Department of Mathematics and Physics, Kochi University, Kochi 780-8520, Japan\\
$^{2}$  Department of Electronic and Physical Systems, Waseda University, Tokyo, 169-8555, Japan\\
$^{3}$  Center for Optical Quantum Technologies, Department of Physics, University of Hamburg, Luruper Chaussee 149, 22761 Hamburg, Germany
}
\begin{abstract}
The notion of a polaron, originally introduced in the context of electrons in ionic lattices, helps us to understand how a quantum impurity behaves when being immersed in and interacting with a many-body background. 
We discuss the impact of the impurities on the medium particles by considering feedback effects from polarons that can be realized in ultracold quantum gas experiments. 
In particular, we 
exemplify the modifications of the medium  in the presence of {}{either} Fermi or Bose polarons. 
Regarding Fermi polarons we present a corresponding many-body diagrammatic approach operating at finite temperatures and discuss how mediated two- and three-body interactions are implemented within this framework. 
Utilizing this approach, we analyze the behavior of the spectral function of Fermi polarons at finite temperature by varying impurity-medium interactions as well as spatial dimensions from three to one.
Interestingly, we reveal that the spectral function of the medium atoms could be a useful quantity for analyzing the transition/crossover from attractive polarons to molecules in three-dimensions.
As for the Bose polaron, we showcase the depletion of the background Bose-Einstein condensate in the vicinity of the impurity atom. 
Such spatial modulations would be important for future investigations regarding the quantification of interpolaron correlations in Bose polaron problems.
\end{abstract}
\maketitle

\section{Introduction}
\label{sec1}
The quantum many-body problem, which is one of the central issues of modern physics, is encountered in various research fields such as condensed matter and nuclear physics. 
The major obstacle that prevents their adequate description stems from {}{the presence of} many degrees-of-freedom as well as strong correlations.
The polaron concept, which was originally proposed by S. I. Pekar and L. Landau~\cite{Pekar1946,Landau1948} to characterize electron properties in crystals, provides a useful playground for understanding related nontrivial many-body aspects of quantum matter and interactions.
For instance, a~key advantage of the polaron picture is that, under~specific circumstances, it enables the reduction of a complicated many-body problem to an effective single-particle or a few-body one with renormalized parameters. 
{}{In the last decade, the~polaron concept has been intensively studied for two-component ultracold mixtures, where a minority component is embedded in a majority one (host) and becomes dressed by the low-energy excitations of the latter forming a polaron. Indeed,}
ultracold atoms, owing to the excellent controllability of the {}{involved system} parameters, are utilized to quantitatively determine polaron properties, as~has been demonstrated in a variety of relevant experimental efforts. 
{}{These include, for~instance, the~measurement of the relevant quasiparticle excitation spectra~\cite{Nascimbene2009,Schirotzek2009,Kohstall2011,Koschorreck2012,Hohmann2015,Jorgensen2016,Hu2016,Scazza2017,Oppong2019,Adlong2020}, monitoring the quantum dynamics of impurities~\cite{Catani2012,Scelle2013}, the~observation of a phononic Lamb shift~\cite{Rentrop2016}, the~estimation of relevant thermodynamic quantities~\cite{Yan2019,Ness2020}, the~identification of medium induced interactions~\cite{DeSalvo2019,Edri2020}, and~polariton properties~\cite{Peyronel2012,Ningyuan2016,Thompson2017}.}
\par
Polarons basically appear in two different types, namely, Fermi and Bose polarons where the impurity atoms are immersed in a Fermi sea and a Bose-Einstein condensate (BEC) respectively. 
Both cases are experimentally realizable by employing a mixture of atoms residing in different hyperfine states or using distinct isotopes. 
The impurity-medium interaction strength can be flexibly adjusted with the aid of Feshbach resonances~\cite{Chin2010}, and~as such strong interactions between the impurity and the majority atoms can be achieved.
Due to this non-zero interaction, the~impurities are subsequently dressed by the elementary excitations of their background atoms, leading to a quasi-particle state that is called the polaron. 
In that light, the~polaron and more generally the quasiparticle generation is inherently related to the build-up of strong entanglement among the impurities and their background medium~\cite{Massignan2014,Schmidt2018,Mistakidis2019a}. 
Moreover, since various situations such as mass-imbalanced~\cite{Kohstall2011}, low-dimensional~\cite{Koschorreck2012}, and~multi-orbital~\cite{Oppong2019} ultracold settings can be realized, atomic polarons can also be expected to be quantum simulators of quasiparticle states in nuclear physics~\cite{Kutschera1993,Forbes2014,TajimaANM2019,Nakano2020,Vidana2021}. 
Recently, a~Rydberg Fermi polaron has also been discussed theoretically~\cite{Sous2020}.
\par
The single-particle character of polarons has been intensively investigated theoretically in the past few years by using different approaches~\cite{Chevy2006,Combescot2007,ProkofevPRB2008,ProkofevPRBR2008,Combescot2008,Bruun2010,Cui2010,Trefzger2012,Mathy2011,Schmidt2011,Kamikado2017,Vlietinck2013,Kroiss2015,Goulko2016,Houcke2020,LiuPRA2020,LiuPRL2020,Baarsma2012} ranging from variational treatments~\cite{Chevy2006,Combescot2007,Combescot2008,Cui2010} to diagrammatic Monte-Carlo simulations~\cite{ProkofevPRB2008,ProkofevPRBR2008,Vlietinck2013,Kroiss2015,Goulko2016,Houcke2020}. 
Interestingly, a~multitude of experimental observations regarding polaronic excitations have been well described based on theoretical frameworks relying on the single-polaron ansatz~\cite{Nascimbene2009,Schirotzek2009,Scazza2017}.
However, it is still a challenging problem and highly unexplored topic how many polaron systems behave, especially during their nonequilibrium dynamics. While the single-polaron analysis clarifies the mechanism of polaron formation via the dressing from the surrounding majority cloud, the~many-polaron study is dedicated to the question of how polarons interact with each other through the exchange of the excitations of their host. 
Therefore, the~background medium plays a crucial role in understanding many-polaron physics.
In this sense, the~concept of induced interpolaron interactions has attracted a tremendous attention~\cite{Pilati2008,Mora2010,Giraud2012,Hu2018,Tajima2018,Tajima2019,Tajimahydro,Takahashi2020,MistakidisPPS,Mukherjee2020,Mistakidisinduced}.
For instance, in~recent experiments, the~sizable shift of the effective scattering length due to the fermion-mediated interaction has been observed in Fermi polaron systems~\cite{DeSalvo2019,Edri2020}.
The corresponding impact on the medium atoms due to the presence of strong impurity-bath correlations is under active investigation~\cite{Tajima2018}. 
In the case of Bose polarons~\cite{Catani2012, Scelle2013, Hohmann2015, Compagno2017, Jorgensen2016, Hu2016, Rentrop2016,Sous2017a,Nakano2017, Nakano2019,Ardila2019,Ardila2020}, the~influence of the impurities on their environment (BEC) is more pronounced when compared to Fermi polarons due to the absence of the Pauli blocking effect.
Characteristic examples, here, constitute the self-localization~\cite{Cucchietti2006, Sacha2006, Kalas2006, Bruderer2008, Boudjemaa2014,Sous2017b} and temporal orthogonality catastrophe~\cite{Mistakidis2019a} phenomena as well as complex tunneling~\cite{TheelDW,KeilerDopped,Sieglexp,Caitransp} and emergent relaxation processes~\cite{MistakidisDiss,MistakidisPPS}. 
They originate from the presence of the impurity which imprints significant deformations to its environment when the interaction between the subsystems is finite.  
\par
In this work, we first provide a discussion on the role of the background atoms in many-polaron problems that are tractable in ultracold atom settings. 
Particularly, we present diagrammatic approaches to Fermi polaron systems and elaborate on how mediated two- and three-body interpolation interactions are consistently taken into account within these frameworks~\cite{Tajima2018,Tajima2019}. 
Importantly, a~comparison of the Fermi polaron excitation spectral function in three dimensions (3D) and at finite temperatures is performed among different variants of the diagrammatic $T$-matrix approach. 
Namely, the~usual $T$-matrix approach (TMA) which is based on the self-energy including the repeated particle-particle scattering processes consisting of bare propagators~\cite{Strinati2018,Ohashi2020}, the~extended $T$-matrix approach (ETMA) where the bare propagator in the self-energy is partially replaced~\cite{Kashimura2012,Tajima2017,Horikoshi2017}, and~the self-consistent $T$-matrix approach where all the propagators in the self-energy consist of dressed ones~\cite{Haussmann1993,Haussmann2007} are employed.
We reveal how medium-induced interactions are involved in these approaches and examine their effects in mass-balanced Fermi polaron settings realized, e.g.,~in $^6$Li atomic mixtures. 
Subsequently, we discuss the polaron excitation spectrum in two (2D) and one (1D) spatial dimensions. 
The behavior of the spectral function of the host and the impurities at strong impurity-medium interactions is exemplified. 
Finally, the~real-space Bogoliubov approach to Bose polarons is reviewed. 
The latter allows us to unveil the condensate deformation due to the presence of the impurity and appreciate the resultant quantum fluctuations  
~\cite{Takahashi2019}. 
We argue that the degree of the quantum depletion of the condensate decreases (increases) for repulsive (attractive) impurity-medium interactions, a~result that is associated with the deformation of its density distribution. 
This is in contrast to homogeneous setups where the depletion increases independently of the sign of the interaction.
\par
This work is organized as follows.
In Section~\ref{sec2},
we present the model Hamiltonian describing ultracold Fermi polarons in three dimensions.
For the Fermi polaron, we consider uniform systems and develop the concept of the diagrammatic $T$-matrix approximation. 
After explaining the ingredients of the diagrammatic approaches in some detail, we clarify how mediated two- and three-body interactions are incorporated in these approaches. 
The behavior of the resultant polaron spectral function at finite temperatures and impurity concentrations in three- two- and one-dimensions is discussed. 
In Section~\ref{sec3}, we utilize the real-space mean-field formulation for Bose polarons and expose the presence of quantum depletion for the three-dimensional trapped Bose polaron at zero temperature.
In \mbox{Section~\ref{sec4}}, we summarize our results and provide future perspectives.
For convenience, in~what follows, we use $k_{\rm B}=\hbar=1$. 

\section{Fermi~Polarons}
\label{sec2}

\subsection{$T$-Matrix Approach to Fermi Polaron~Problems}
\label{sec2-1}

Here we explain the concept of many-body diagrammatic approaches to Fermi polarons, namely, settings referring to the situation where fermionic impurity atoms are immersed in a uniform Fermi gas. 
Since such a two-component Fermi mixture mimics spin-$1/2$ electrons,
we denote the bath component as ${\sigma=\rm B}=\up$ and the impurity one by $\sigma={\rm I}=\down$. Note that these are standard conventions without loss of generality.
The model Hamiltonian describing this system reads
\begin{align}
\label{eq:H}
H=\sum_{\bm{p},\sigma}\xi_{\bm{p},\sigma}c_{\bm{p},\sigma}^\dag c_{\bm{p},\sigma} 
+g\sum_{\bm{p},\bm{p}',\bm{q}}
c_{\bm{p}+\bm{q}/2,{\up}}^\dag
c_{-\bm{p}+\bm{q}/2,{\down}}^\dag c_{-\bm{p}'+\bm{q}/2,{\down}}
c_{\bm{p}'+\bm{q}/2,{\up}},
\end{align}
where $\xi_{\bm{p},\sigma}=p^2/(2m_\sigma)-\mu_\sigma$ is the kinetic energy minus the chemical potential $\mu_\sigma$, and~$m_\sigma$ is the atomic mass of the $\sigma$ component. 
The parameters $c_{\bm{p},\sigma}$ and $c_{\bm{p},\sigma}^\dag$ refer to the annihilation and creation operators of a $\sigma$ component fermion, respectively, possessing momentum $\bm{p}$.
\par
We measure the effective coupling constant $g$ of the contact-type interaction between two different component fermions by using the low-energy scattering parameter, namely, the~scattering length $a$.
In 3D, it is known~\cite{PethickSmith} 
that the coupling constant $g_{\rm 3D}$ and the scattering length $a$ are related via
\begin{align}
\frac{m_{\rm r}}{2\pi a}=\frac{1}{g_{\rm 3D}}+\frac{m_{\rm r}\Lambda}{\pi^2},
\end{align}
with $m_{\rm r}^{-1}=m_{\up}^{-1}+m_{\down}^{-1}$ being the reduced mass. 
In this expression, the~momentum cutoff $\Lambda$ is introduced to avoid an ultraviolet divergence in the momentum summation of the Lippmann--Schwinger equation expressed in momentum space. 
{This allows us to achieve the effective short-range interaction of finite range $r_{\rm e}\propto 1/\Lambda$}. 
Similarly, the~relevant relations in 2D and 1D read~\cite{Morgan2002}
\begin{align}
\label{eq:a}
a_{\rm 2D}=\frac{1}{\Lambda}e^{-\frac{\pi}{m_{\rm r}g_{\rm 2D}}},~~~{\rm and}~~~ 
a_{\rm 1D}=\frac{1}{m_{\rm r}g_{\rm 1D}},
\end{align}
respectively, where $g_{\rm 2D}$ and $g_{\rm 1D}$ are the coupling constants in 2D and~1D.
 
First, we introduce a thermal single-particle Green's function~\cite{Fetter2003}
\begin{align}
G_{\sigma}(\bm{p},i\omega_n)= \frac{1}{i\omega_n-\xi_{\bm{p},\sigma}-\Sigma_{\sigma}(\bm{p},i\omega_n)},   
\end{align}
where $\omega_n=(2n+1)\pi T$ is the fermion Matsubara frequency introduced within the finite-temperature $T$ formalism and $n\in \mathbb{Z}$~\cite{Fetter2003}. 
The effect of the impurity-medium interaction is taken into account in the self-energy $\Sigma_{\sigma}(\bm{p},i\omega_n)$. 
The excitation spectrum $A_{\down}(\bm{p},\omega)$ of a Fermi polaron can be obtained via the retarded Green's function $G_{\down}^{\rm R}(\bm{p},\omega)=G_{\down}(\bm{p},i\omega_n\rightarrow \omega+i\delta)$ (where $\delta$ is a positive infinitesimal) through analytic continuation~\cite{Fetter2003}. 
In particular, it can be shown that
\begin{align}
A_{\down}(\bm{p},\omega)=-\frac{1}{\pi}{\rm Im}G_{\down}^{\rm R}(\bm{p},\omega).    
\end{align}

Experimentally, this quantity can be monitored by using a radio-frequency (rf) spectroscopy scheme where the atoms are transferred from their thermal equilibrium state to a specific spin state which interacts with the medium~\cite{Torma2015}. 
Indeed, the~reverse rf response $I_{\rm r}(\omega)$~\cite{Scazza2017} and the ejection one $I_{\rm e}(\omega)$~\cite{Yan2019} are given by
\begin{align}
\label{eq:Ir}
I_{\rm r}(\omega)=2\pi\Omega_{\rm Rabi}^2\sum_{\bm{p}}f(\xi_{\bm{p},{\rm i}})A_{\down}(\bm{p},\omega+\xi_{\bm{p},\down})
\end{align}
and
\begin{align}
\label{eq:Ie}
I_{\rm e}(\omega)=2\pi\Omega_{\rm Rabi}^2\sum_{\bm{p}}f(\xi_{\bm{p},\down}-\omega)A_{\down}(\bm{p},\xi_{\bm{p},\down}-\omega),     
\end{align}
respectively. 
Here, $\xi_{\bm{p},{\rm i}}$ represents the kinetic energy of the initial state in the reverse rf scheme. 
In Equations~(\ref{eq:Ir}) and (\ref{eq:Ie}), $\Omega_{\rm Rabi}$ is the Rabi frequency. 
\par
Importantly, 
the self-energy $\Sigma_{\up}(\bm{p},i\omega_n)$ of the background plays an important role in describing the mediated interpolaron interactions. 
This fact will be evinced below and it is achieved by expanding $\Sigma_{\up}(\bm{p},i\omega_n)$ with respect to $G_{\sigma}$ and $G_{\sigma}^0$. 
The~chemical potentials $\mu_\sigma$ are kept fixed by imposing the particle number conservation condition obeying
\begin{align}
\label{eq:N}
N_{\sigma}=T\sum_{\bm{p},i\omega_n}G_\sigma(\bm{p},i\omega_n). 
\end{align}

{Moreover, in~the remainder of this work, we define the impurity concentration as~follows}
\begin{align}
x=\frac{N_\down}{N_\up}.
\end{align}
\par
\begin{figure}[t]
\includegraphics[width=12cm]{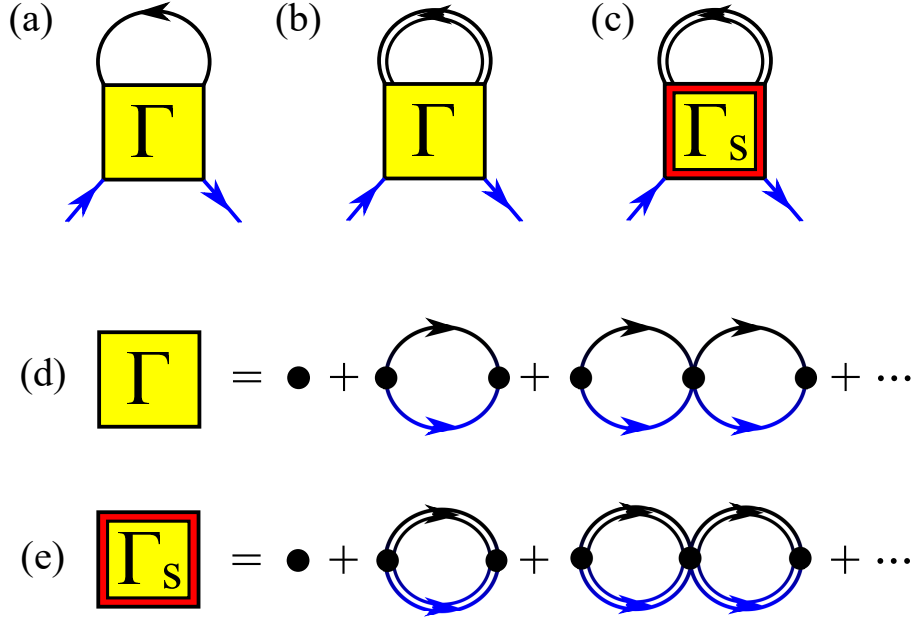}
\caption{Feynman diagrams for (\textbf{a}) the $T$-matrix approach (TMA), (\textbf{b}) the extended $T$-matrix approach (ETMA), and~(\textbf{c}) the self-consistent $T$-matrix approach (SCTMA). 
$\Gamma$ and $\Gamma_{\rm S}$ are the many-body $T$-matrices, whose perturbative expansions are shown schematically in (\textbf{d},\textbf{e}), consisting of bare and dressed propagators $G_{\sigma}^0$ and $G_{\sigma}$, respectively. 
While in TMA, all the lines in the self-energy (\textbf{a})~consist of $G_{\sigma}^0$, they are replaced with $G_{\sigma}$ partially (upper loop of (\textbf{b})) in ETMA and fully in SCTMA (\textbf{c}) (see also (\textbf{e}) where $G_{\sigma}^0$ is replaced by $G_\sigma$ compared to (\textbf{d})), respectively. 
}
\label{fig1}
\end{figure}

Additionally, within~the TMA~\cite{Combescot2007,Hu2018} the self- energy $\Sigma_{\sigma}(\bm{p},i\omega_n)$ of the $\sigma$ component~reads
\begin{align}
\label{eq:sigma}
\Sigma_{\sigma}(\bm{p},i\omega_n)=T\sum_{\bm{q},i\nu_\ell}\Gamma(\bm{q},i\nu_\ell)G_{-\sigma}^0(\bm{q}-\bm{p},i\nu_\ell-i\omega_n),    
\end{align}
where $\Gamma(\bm{q},i\nu_\ell)$ is the many-body $T$-matrix, as~diagrammatically shown in Figure~\ref{fig1}a, with~the boson Matsubara frequency $i\nu_{\ell}=2\ell \pi T$ ($\ell\in \mathbb{Z}$).
Here, $G_{\sigma}^0(\bm{p},i\omega_n)=(i\omega_n-\xi_{\bm{p},\sigma})^{-1}$ is the bare thermal single-particle Green's function. 
Furthermore, by~adopting a ladder approximation illustrated in Figure~\ref{fig1}d, the~$T$-matrix
$\Gamma(\bm{q},i\nu_\ell)$ is given by
\begin{align}
\Gamma(\bm{q},i\nu_\ell)=\frac{g}{1+g\Pi(\bm{q},i\nu_\ell)},
\end{align}
where
\begin{align}
\label{eq:pi}
\Pi(\bm{q},i\nu_\ell)=T\sum_{\bm{p},i\omega_n}G_{\up}^0(\bm{p}+\bm{q},i\omega_n+i\nu_\ell)G_{\down}^0(-\bm{p},-i\omega_n)   
\end{align}
is the lowest-order particle-particle bubble. 
The latter describes a virtual particle-particle scattering process associated with the impurity-medium interaction $g$ which is replaced by $g_{\rm 3D}$, $g_{\rm 2D}$, and~$g_{\rm 1D}$ in 3D, 2D, and~1D, respectively.
Note that in Equation~(\ref{eq:sigma}) the impurity-impurity interaction is not taken into account. 
\par
The extended {\it T}-matrix approach (ETMA)~\cite{Tajima2018} constitutes an improved approximation that allows us to take the induced polaron-polaron interactions into account in a self-consistent way.
In this method, as~depicted in Figure~\ref{fig1}b we include higher-order correlations by replacing the bare Green function $G^0$ in Equation~(\ref{eq:sigma}) with the dressed one $G_\sigma$.
Namely
\begin{align}
\label{eq:sigE}
\Sigma_{\sigma}^{\rm E}(\bm{p},i\omega_n)=T\sum_{\bm{q},i\nu_\ell}\Gamma(\bm{q},i\nu_\ell)G_{-\sigma}(\bm{q}-\bm{p},i\nu_\ell-i\omega_n).
\end{align}

Importantly, the~TMA and ETMA approaches are equivalent to each other in the single-polaron limit i.e.,~$x\rightarrow 0$, where the self-energy of the fermionic medium $\Sigma_{\up}^{\rm E}$ (capturing the difference between $G_{\up}^0$ and $G_{\up}$ in Equations~(\ref{eq:sigma}) and (\ref{eq:sigE}), respectively) is negligible.
Additionally, at~zero temperature, these two treatments coincide with the variational ansatz proposed by F. Chevy~{\cite{Chevy2006}}. 
Recall that $\mu_{\up}=E_{\rm F}$ and $\mu_{\down}=E_{\rm P}^{\rm (a)}$ at $T=0$ and $x\rightarrow 0$, where $E_{\rm F}=p_{\rm F}^2/(2m_\up)$ denotes the Fermi energy of the majority component atoms while $E_{\rm P}^{\rm (a)}$ corresponds to the attractive polaron energy.
\par
Proceeding one step further, it is possible to construct the so-called self-consistent $T$-matrix approach (SCTMA)~\cite{Frank2018,Pini2019,Tajima2019} which deploys the many-body $T$-matrix $\Gamma_{\rm S}$ composed of dressed propagators as schematically shown in Figure~\ref{fig1}e.
In particular, the~corresponding $T$-matrix is given by
\begin{align}
\Gamma_{\rm S}(\bm{q},i\nu_\ell)=\frac{g}{1+g\Pi_{\rm S}(\bm{q},i\nu_\ell)},
\end{align}
where
\begin{align}
\Pi_{\rm S}(\bm{q},i\nu_\ell)=T\sum_{\bm{p},i\omega_n}G_{\up}(\bm{p}+\bm{q},i\omega_n+i\nu_\ell)G_{\down}(-\bm{p},-i\omega_n),
\end{align}
which describes a scattering process denoted by $G_{\uparrow }$ and $G_{\downarrow}$, of~the dressed medium atoms with the impurities and the dressed ones (polarons), respectively. 
This is in contrast to Equation~(\ref{eq:pi}) obtained in ETMA and consisting of $G_{\sigma}^0$ which represents the impurity-medium scattering process of only the bare atoms.  
Using this $T$-matrix, we can express the SCTMA self-energy $\Sigma_{\sigma}^{\rm S}$ (see also Figure~\ref{fig1}c) as
\begin{align}
\Sigma_{\sigma}^{\rm S}(\bm{p},i\omega_n)=T\sum_{\bm{q},i\nu_\ell}\Gamma_{\rm S}(\bm{q},i\nu_\ell)G_{-\sigma}(\bm{q}-\bm{p},i\nu_\ell-i\omega_n).
\end{align}
  
\par
We note that within the ETMA, the~impurity self-energy $\Sigma_{\down}^{\rm E}$ (Equation~(11)) can be rewritten~as
\begin{align}
\Sigma_{\down}^{\rm E}(\bm{p},i\omega_n)&=T\sum_{\bm{q},i\nu_\ell}\Gamma(\bm{q},i\nu_\ell)\left[G_{\up}^0(\bm{q}-\bm{p},i\nu_\ell-i\omega_n)\right.\cr
&+\left.G_{\up}^0(\bm{q}-\bm{p},i\nu_\ell-i\omega_n)\Sigma_{\up}(\bm{q}-\bm{p},i\nu_\ell-i\omega_n)G_{\up}(\bm{q}-\bm{p},i\nu_\ell-i\omega_n)\right]\cr
&\equiv \Sigma_{\down}(\bm{p},i\omega_n)+\delta\Sigma_{\down}(\bm{p},i\omega_n),
\end{align}
with the higher-order correction $\delta\Sigma_{\down}(\bm{p},i\omega_n)$ beyond the TMA being 
\begin{align}
\delta\Sigma_{\down}(\bm{p},i\omega_n)&=T^2\sum_{\bm{q},\bm{q}',i\nu_\ell,i\nu_{\ell'}}\Gamma(\bm{q},i\nu_\ell)\Gamma(\bm{q}',i\nu_{\ell'})G_{\up}^0(\bm{q}-\bm{p},i\nu_\ell-i\omega_n)G_{\up}(\bm{q}-\bm{p},i\nu_\ell-i\omega_n)\cr
&\quad\quad\quad\quad\quad
\times G_{\down}(\bm{q}'-\bm{q}+\bm{p},i\nu_{\ell'}-i\nu_{\ell}+i\omega_n)\cr
&\equiv T\sum_{\bm{p}',i\omega_{n'}} V_{\rm eff}^{(2)}(\bm{p},i\omega_n,\bm{p}',i\omega_{n'};\bm{p},i\omega_n,\bm{p}',i\omega_{n'})G_{\down}(\bm{p}',i\omega_{n'}).
\end{align}

\begin{figure}[t]
\includegraphics[width=11cm]{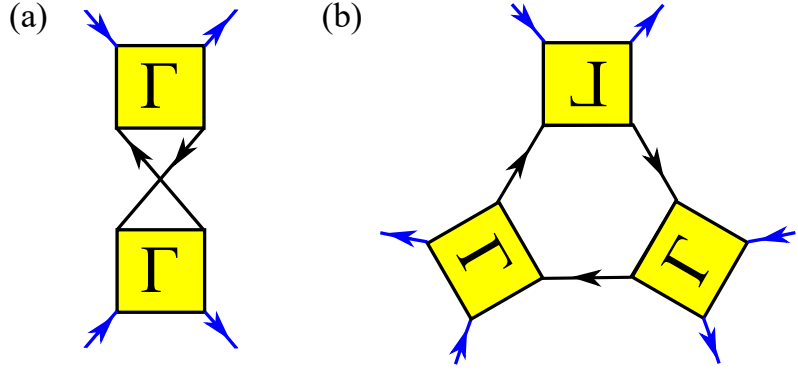}
\caption{Feynman diagrams for induced (\textbf{a}) two- and
(\textbf{b}) three-body interactions $V_{\rm eff}^{(2,3)}$ among polarons. 
The arrows represent the direction of momentum and energy transfer in each propagator.}
\label{fig2}
\end{figure}

In this expression, $V_{\rm eff}^{(2)}(\bm{p}_1,i\omega_{n_1},\bm{p}_{2},i\omega_{n_2};\bm{p}_1',i\omega_{n_1'},\bm{p}_2',i\omega_{n_2'})$ represents the induced impurity-impurity interaction (diagrammatically shown in Figure~\ref{fig2}a) with incoming and outgoing momenta and frequencies $\{\bm{p}_i,i\omega_{n_i}\}$ and $\{\bm{p}_i',i\omega_{n_i'}\}$, respectively, where $i=1,2$. 
It reads
\begin{align}
&V_{\rm eff}^{(2)}(\bm{p}_1,i\omega_{n_1},\bm{p}_{2},i\omega_{n_2};\bm{p}_1',i\omega_{n_1'},\bm{p}_2',i\omega_{n_2'})=
\delta_{\bm{p}_1+\bm{p}_2,\bm{p}_1'+\bm{p}_2'}\delta_{n_1+n_2,n_1'+n_2'}\cr
&\times T\sum_{\bm{q},i\nu_{\ell}}
\Gamma(\bm{q},i\nu_\ell)\Gamma(\bm{q}+\bm{p}_2-\bm{p}_1',i\nu_{\ell}+i\omega_{n_2}-i\omega_{n_1'})
G_{\up}^0(\bm{q}-\bm{p}_1,i\nu_\ell-i\omega_{n_1})G_{\up}^0(\bm{q}-\bm{p}_1',i\nu_\ell-i\omega_{n_1'}).
\end{align}

Here, $\delta_{i,j}$ is the Kronecker delta imposing the energy and momentum conservation in the two-body scattering. 
\par
The self-energy $\Sigma_{\down}^{\rm S}$ of the impurities within the SCTMA involves a contribution of induced three-impurity correlations due to the dressed pair propagator $\Sigma_{\down}^{\rm S}$.
The latter can again be decomposed as
\begin{align}
\Sigma_{\down}^{\rm S}(\bm{p},i\omega_n)\equiv\Sigma_{\down}^{\rm E}(\bm{p},i\omega_n)+\delta\Sigma_{\down}'(\bm{p},i\omega_n),
\end{align}
where
\begin{align}
\delta\Sigma_{\down}'(\bm{p},i\omega_n)&=T\sum_{\bm{q},i\nu_\ell}\left[\Gamma_{\rm S}(\bm{q},i\nu_\ell)-\Gamma(\bm{q},i\nu_\ell)\right]G_{\up}(\bm{q}-\bm{p},i\nu_\ell-i\omega_n)\cr
&=T\sum_{\bm{q},i\nu_\ell}\Gamma_{\rm S}(\bm{q},i\nu_\ell)\Gamma(\bm{q},i\nu_\ell)\Phi(\bm{q},i\nu_\ell)G_{\up}(\bm{q}-\bm{p},i\nu_\ell-i\omega_n).
\end{align}

Here we defined
\begin{align}
\label{eq:phi}
\Phi(\bm{q},i\nu_\ell)&=\Pi_{\rm S}(\bm{q},i\nu_\ell)-\Pi(\bm{q},i\nu_\ell)\cr
&=T\sum_{\bm{p},i\omega_n}\left[G_{\up}(\bm{p}+\bm{q},i\omega_n+i\nu_\ell)G_{\down}(-\bm{p},-i\omega_n)-G_{\up}^0(\bm{p}+\bm{q},i\omega_n+i\nu_\ell)G_{\down}^0(-\bm{p},-i\omega_n)\right]\cr
&\simeq T\sum_{\bm{p},i\omega_n}\left[G_{\up}^0(\bm{p}+\bm{q},i\omega_n+i\nu_\ell)\right]^2\Sigma_{\up}^{\rm S}(\bm{p}+\bm{q},i\omega_n+i\nu_\ell)G_{\down}^0(-\bm{p},-i\omega_n),
\end{align} 
which represents the difference between the $\Pi$ and $\Pi_{\rm S}$, namely, the~medium-impurity and the medium-polaron propagators.
In the last line of Equation~(\ref{eq:phi}),
we assumed that $G_{\up}\simeq G_{\up}^0$ and $\Sigma_{\down}^{\rm S}\simeq 0$.
Thus, one can find a three-body correlation effect beyond the ETMA as shown in Figure~\ref{fig2}b and captured by
\begin{align}
\delta\Sigma_{\down}'(\bm{p},i\omega_n)\simeq & T\sum_{\bm{p}',i\omega_{n'}}V_{\rm eff}^{(3)}(\bm{p},i\omega_{n},\bm{p}',i\omega_{n'},\bm{p}'',i\omega_{n''};\bm{p}',i\omega_{n'},\bm{p},i\omega_{n},\bm{p}'',i\omega_{n''})\cr
&\times G_{\down}(\bm{p}',i\omega_{n'})G_{\down}^0(\bm{p}'',i\omega_{n''}),
\end{align}
where $V_{\rm eff}^{(3)}(\bm{p}_1,i\omega_{n_1},\bm{p}_2,i\omega_{n_2},\bm{p}_3,i\omega_{n_3};\bm{p}_1',i\omega_{n_1'},\bm{p}_2',i\omega_{n_2'},\bm{p}_3',i\omega_{n_3'})$ is the induced three-polaron interaction term.
Its explicit form reads
\begin{align}
&V_{\rm eff}^{(3)}(\bm{p}_1,i\omega_{n_1},\bm{p}_2,i\omega_{n_2},\bm{p}_3,i\omega_{n_3};\bm{p}_1',i\omega_{n_1'},\bm{p}_2',i\omega_{n_2'},\bm{p}_3',i\omega_{n_3'})
=\delta_{\bm{p}_1+\bm{p}_2+\bm{p}_3,\bm{p}_1'+\bm{p}_2'+\bm{p}_3'}
\delta_{n_1+n_2+n_3,n_1'+n_2'+n_3'}\cr 
&\quad\times T\sum_{\bm{q},i\nu_\ell}\Gamma(\bm{q},i\nu_\ell)\Gamma(\bm{q}+\bm{p}_3-\bm{p}_1',i\nu_{\ell}+i\omega_{n_3}-i\omega_{n_1'})\Gamma(\bm{q}+\bm{p}_2'-\bm{p}_1,i\nu_{\ell}+i\omega_{n_2'}-i\omega_{n_1})\cr
&\quad\times G_{\up}^0(\bm{q}-\bm{p}_1',i\nu_\ell-i\omega_{n_1'})
G_{\up}^0(\bm{q}-\bm{p}_1,i\nu_\ell-i\omega_{n_1})
G_{\up}^0(\bm{q}-\bm{p}_1'+\bm{p}_3-\bm{p}_{3}',i\nu_\ell-i\omega_{n_1'}+i\omega_{n_3}-i\omega_{n_3'}).
\end{align}

From the above discussion, it becomes evident how the medium-induced two-body and three-body interpolaron interactions are included in the ETMA and the SCTMA treatments.
Recall that in the TMA the interpolaron interaction is not taken into account.
Even so, observables such as thermodynamic quantities (e.g., particle number density) and spectral functions obtained via rf spectroscopy can in principle provide indications of the effect of interpolaron interactions 
through $\Sigma_{\sigma}(\bm{p},i\omega_n)$.

\subsection{Spectral Response of Fermi~Polarons}

In the following, we shall present and discuss the behavior of the spectral function of Fermi polarons for temperatures ranging from zero to the Fermi temperature of the majority component as well as for different spatial dimensions from three to one. 
For simplicity, we consider a mass-balanced fermionic mixture i.e.,~$m_\up=m_\down\equiv m$. 
The latter is experimentally relevant for instance by considering two different hyperfine states, e.g.,~$|F=1/2,m_{\rm F}=+1/2\rangle$ and $|F=3/2,m_{\rm F}=-3/2\rangle$ of $^6$Li. In~this notation, $F$ and $m_{\rm F}$ are the total angular momentum and its projection, respectively, of~the specific hyperfine state~\cite{Scazza2017}  at thermal equilibrium. 

\subsubsection{Three-Dimensional~Case}

The resultant spectral function $A_{\sigma}(\bm{p}=0,\omega)$ of the fermionic medium ($\sigma=\up$) and the impurities ($\sigma=\down$) is depicted in Figure~\ref{fig3} 
as a function of the single-particle energy $\omega$.  
Here, we consider a temperature $T=0.3T_{\rm F}$, impurity concentration $x=0.1$, and~impurity-medium interaction at unitarity, i.e.,~$(p_{\rm F}a)^{-1}=0$. 
The Fermi temperature is $T_{\rm F}=p_{\rm F}^2/(2m_\up)$ and the Fermi momentum $p_{\rm F}$.
Evidently, the~spectral function of the majority component (Figure~\ref{fig3}a) exhibits a peak around $\omega+\mu_\up=0$ in all three diagrammatic approaches introduced in Section~\ref{sec2}. 
The sharp peak around $\omega+\mu_\up=0$ corresponds to the spectrum of the bare medium atoms given by $A(\bm{p},\omega)=\delta(\omega-\xi_{\bm{p},\up})$ at $\bm{p}=0$. 
This indicates that the imprint of the impurity-medium interaction on the fermionic host is negligible for such small impurity concentrations $x=0.1$; see also {Figure}
~\ref{fig_addfor3d} and the discussion below. 
Indeed, the~renormalization of $\mu_\up$ (which essentially evinces the backaction on the majority atoms from the impurities) in the ETMA at unitarity is proportional to $x$~\cite{Tajima2018} and in particular
\begin{align}
\label{eq:muup}
\frac{\mu_{\up}}{E_{\rm F}}=1-0.526x.    
\end{align}

It can be shown that in the weak-coupling limit,
this shift is given by the Hartree correction
$\Sigma_{\up}^{\rm H}=\frac{4\pi a}{m}N_{\down}$~\cite{Fetter2003}.
However, at~the unitarity limit presented in Figure~\ref{fig3},
such a weak-coupling approximation cannot be applied and therefore the factor $0.526$ in \mbox{Equation~(\ref{eq:muup})} originates from the existence of strong correlations between the majority and the minority component atoms.

{\begin{figure}[t]
\includegraphics[width=13cm]{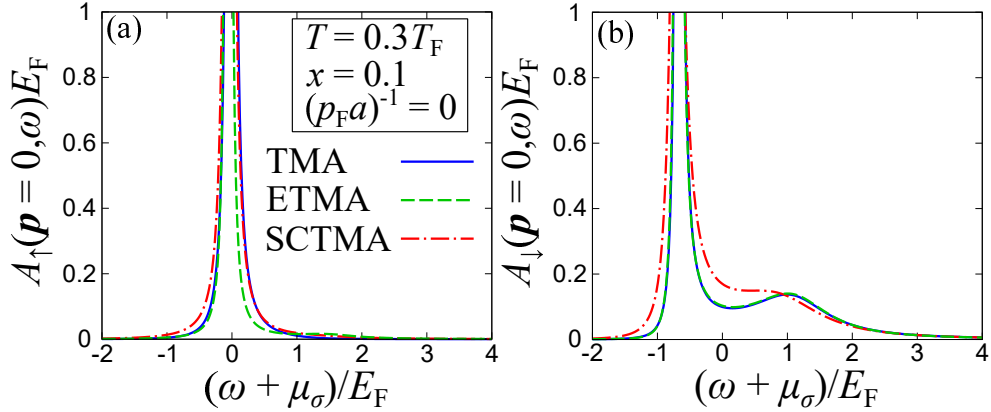}
\caption{Zero
-momentum spectral functions $A_{\sigma}(\bm{p}=\bm{0},\omega)$ of (\textbf{a}) the majority (medium) and (\textbf{b})~the minority (impurities) fermions for varying energy $\omega$ at unitarity, $(p_{\rm F}a)^{-1}=0$.
We consider a temperature $T=0.3T_{\rm F}$ and an impurity concentration $x=0.1$.
The solid, dashed, and~dash-dotted lines represent
the results of the TMA, ETMA, and~SCTMA approaches respectively. While \mbox{$A_{\uparrow}(\bm{p}=\bm{0},\omega)$} is almost the same among the three approaches, $A_{\down}(\bm{p}=\bm{0},\omega)$ within the SCTMA experiences a sizable difference compared to the response obtained in the TMA and the ETMA~approaches. }
\label{fig3}
\end{figure}   
\par
The corresponding polaronic excitation spectrum is captured by $A_{\down}(\bm{p}=0,\omega)$ \linebreak {({Figure~\ref{fig3}b}) having a dominant} peak at 
$\omega+\mu_\down=-E_{\rm P}^{\rm (a)}$ where $E_{\rm P}^{\rm (a)}$ is the attractive polaron energy.
Notice here that since this peak is located at negative energies it indicates the formation of an attractive Fermi polaron. 
This observation can be understood from the fact that in the absence of impurity-medium interactions, the~bare-particle pole, namely, the~position of the pole of the bare retarded single-particle Green's function $G_{\down}^{0,{\rm R}}(\bm{p}=\bm{0},\omega)=(\omega+i\delta+\mu_\down)^{-1}$, occurs at $\omega+\mu_\down=0$. 
Moreover, the~attractive polaron energy $E_{\rm P}^{\rm (a)}$ (being of course negative) is defined by the self-energy energy shift as $E_{\rm P}^{\rm (a)}=\Sigma_{\down}(\bm{0},E_{\rm P}^{\rm (a)})$.
Thus, one can regard the deviation of the position of the peak from $\omega+\mu_\down=0$ as the attractive polaron energy $E_{\rm P}^{\rm (a)}$, since it is given by
$A_{\down}(\bm{p}=0,\omega)\sim \delta(\omega+\mu_{\down}-E_{\rm P}^{\rm (a)})$. 
Recall that, in~general, for~finite temperatures $T$ and impurity concentrations $x$, $\mu_{\down}\neq E_{\rm P}^{\rm (a)}$ holds in contrast to the single-polaron limit at $T=0$~\cite{Tajima2018}.
Additionally, a~weak amplitude peak appears in $A_{\down}(\bm{p}=0,\omega)$ at positive energies $\omega\simeq E_{\rm F}$.
It stems from the metastable upper branch 
of the impurities, where excited atoms repulsively interact with each other.
This peak becomes sharper at positive scattering lengths away from~unitarity.
  
Indeed, for~positive scattering lengths, the~quasi-particle excitation called a repulsive Fermi polaron emerges~\cite{Massignan2014}. 
\par
Figure~\ref{figpol}a presents the polaron spectral function $A_{\down}(\bm{p}=\bm{0},\omega)$ with respect to the interaction parameter $(p_{\rm F}a)^{-1}$ obtained within the ETMA method at $T=0.03T_{\rm F}$ and $x=O(10^{-4})$.
From the position of the poles of $G_{\down}^{\rm R}(\bm{p}=\bm{0},\omega)$,
one can extract two kinds of polaron energies, namely, $E_{\rm P}^{\rm (a)}$ and $E_{\rm P}^{\rm (r)}$ corresponding to the attractive and the repulsive polaron energies, respectively. 
The interaction dependence of these energies is provided in Figure~\ref{figpol}b. 
$E_{\rm P}^{\rm (r)}$ approaches the Hartree shift $\Sigma_{\down}^{\rm H}=\frac{4\pi a}{m}N_{\up}$ without the imaginary part of the self-energy (being responsible for the width of the spectra) and finally becomes zero~\cite{Massignan2014}. 
Indeed, the~spectrum in Figure~{\ref{figpol}}a shows that the peak of the repulsive polaron at $\omega+\mu_\down>0$ becomes sharper when increasing $(p_{\rm F}a)^{-1}$, indicating the vanishing imaginary part of the self-energy. 
On the other hand, $E_{\rm P}^{\rm (a)}$ decreases with increasing $(p_{\rm F}a)^{-1}$ as depicted by the position of the low-energy peak (where $\omega+\mu_{\down}<0$) in Figure~\ref{figpol}a. Eventually, the~attractive polaron undergoes the molecule transition as we discuss below.  
Another important issue here is that in 
the strong-coupling regime the attractive polaron undergoes 
the transition to the molecular state with increasing impurity-bath attraction~\cite{Punk2009}.
Although this transition was originally predicted to be of first-order, recent experimental and theoretical studies showed an underlying crossover behavior and coexistence between polaronic and molecular states~\cite{Ness2020}. 
We note that in the case of finite impurity concentrations, a~BEC of molecules can appear at low temperatures; see also Equations~(\ref{eqTC}) and~(\ref{temperature}) below. 
It is also a fact that the interplay among a molecular BEC, thermally excited molecules, and~polarons may occur at finite temperatures~\cite{Cui2020}.
In the calculation of the attractive polaron energy $E_{\rm P}^{{\rm (a)}}$ {for different coupling strengths} (\mbox{Figure~\ref{figpol}b}), however, we do not encounter the molecular BEC transition identified by the Thouless criterion~\cite{Thouless1960}
\begin{align}
\label{eqTC}
1+g\Pi(\bm{q}=\bm{0},i\nu_{\ell}=0)=0.
\end{align}

In particular, in~the strong-coupling limit,
from Equation~(\ref{eqTC}) combined with the particle number conservation (Equation~(\ref{eq:N})) 
the BEC temperature $T_{\rm BEC}$ of molecules satisfies~\cite{Liu2006}
\begin{align}
T_{\rm BEC}\simeq2\pi \left(\frac{x}{12\pi^2\zeta(3/2)}\right)^{\frac{2}{3}}T_{\rm F},\label{temperature}
\end{align}
where $\zeta(3/2)\simeq 2.612$ is the zeta function.
Since we consider a small impurity concentration $x=O(10^{-4})$ here, $T=0.03T_{\rm F}$ is far above $T_{\rm BEC}\propto x^{\frac{2}{3}}$.

{\begin{figure}[t]
\includegraphics[width=13.5cm]{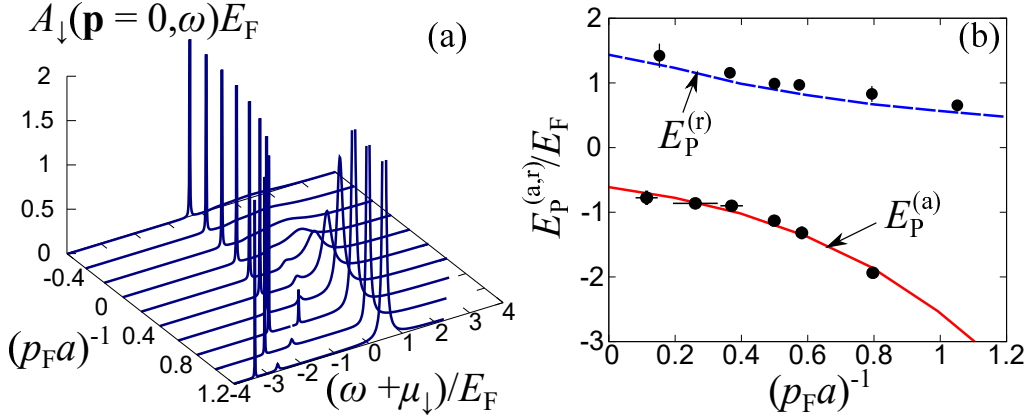}
\caption{(\textbf{a}) Polaron 
 spectral function $A_{\down}(\bm{p}=\bm{0},\omega)$ for several coupling strengths $(p_{\rm F}a)^{-1}$. 
The spectrum is calculated within the ETMA at temperature $T=0.03T_{\rm F}$ and impurity concentration $x=O(10^{-4})$~\cite{Tajima2018}.
Panel (\textbf{b}) represents the attractive and repulsive polaron energies, namely, $E_{\rm P}^{\rm (a)}$ and $E_{\rm P}^{\rm (r)}$, respectively, as~a function of $(p_{\rm F}a)^{-1}$. 
The polaron energies have been extracted from the peak position of $A_{\down}(\bm{p}=\bm{0},\omega)$, that is, the~pole of $G_{\down}^{\rm R}(\bm{p}=\bm{0},\omega)$.
The experimental data of Ref.~\cite{Scazza2017} are plotted in black circles for
direct comparison with the theoretical predictions. 
}
\label{figpol}
\end{figure}

\par
According to the above-description, induced polaron-polaron interactions are mediated by the host atoms, which are taken into account within the ETMA and the SCTMA methods as explicated in Section~\ref{sec2},
are weak in the present mass-balanced fermionic mixture. These finite temperature findings are consistent with previous theoretical works~\cite{Pilati2008,Mora2010,Giraud2012} predicting a spectral shift of the polaron energy $\Delta E=FE_{\rm FG}x$ with $F$ = 0.1$\sim$0.2 at \mbox{$T=0$} (where $E_{\rm FG}$ is the ground-state energy of a non-interacting single-component Fermi gas at $T=0$) as well as the experimental observations of Ref.~\cite{Schirotzek2009}. 
On the other hand, the~presence of induced polaron--polaron interactions in the repulsive polaron scenario cannot be observed experimentally~\cite{Scazza2017}, a~result that is further supported by recent studies based on diagrammatic approaches~\cite{Tajima2018}.
\par
Furthermore, the~spectral deviations between the TMA and the ETMA treatments represent the effect of induced two-body interpolaron interactions in the attractive polaron case. 
However, in~our case there is no sizable shift between the spectral lines predicted in these approaches (Figure~\ref{fig3}b).
Indeed, the~induced two-body energy is estimated to be of the order of $10^{-3}E_{\rm FG}$ at $x=0.1$.
The induced three-body interpolaron interaction, which is responsible for the difference among the ETMA and the SCTMA results, exhibits a sizable effect on the width of the polaron spectra.
{}{We remark that at $T=0.3T_{\rm F}$ and $x=0.1$ (Figure~\ref{fig3}b) although the minority atoms basically obey the Boltzmann statistic, since their temperature is higher than the Fermi degenerate temperature $T_{\rm F,\downarrow}=\frac{(6\pi^2 N_{\downarrow})^{\frac{2}{3}}}{2m}$~\cite{Tajima2018} namely $T=0.3T_{\rm F}\simeq 1.39 T_{\rm F,\downarrow}$, effects of the strong medium-impurity interaction on the polaron spectra are present manifesting for instance as a corresponding broadening.} 
Although the SCTMA treatment tends to overestimate the polaron energy, the~observed full-width-at-half maximum (FWHM) of the rf spectrum given by $2.71 (T/T_{\rm F})^2$~\cite{Yan2019} can be well reproduced by this approach. The~latter gives $2.95(T/T_{\rm F})^2$ whereas the FWHM in ETMA is $1.61(T/T_{\rm F})^2$~\cite{Tajima2019}.
We should also note that the decay rate related to the FWHM for repulsive polarons as extracted using TMA (and simultaneously ETMA) agree quantitatively with the experimental result of Ref.~\cite{Scazza2017}. 
For the attractive polaron, the quantitative agreement between the experiment and these diagrammatic approaches is broken at high temperatures.
For instance, the~recent experiment of Ref.~\cite{Yan2019} showed that the transition from polarons 
to the Boltzmann gas occurs at $T\simeq 0.75T_{\rm F}$~\cite{Yan2019}, while the prediction of the diagrammatic approaches is above $T_{\rm F}$~\cite{Tajima2019}.
{Besides the fact that such polaron decay properties may be related to multi-polaron scattering events leading to many-body dephasing~\cite{Adlong2020}, they are necessary for further detailed polaron investigations at various temperatures and interaction strengths that facilitate the understanding of the underlying physics of the observed polaron-to-Boltzmann-gas transition.}
\par
   
The dependence of the polaron spectra $A_{\down}(\bm{p},\omega)$ on the energy and the momentum of the impurities is illustrated in Figure~\ref{fig4} for $T=0.2T_{\rm F}$, $x=0$, and~$(p_{\rm F}a)^{-1}=0$. To~infer the impact of the multi-polaron correlations on the spectrum we explicitly compare $A_{\down}(\bm{p},\omega)$ between the ETMA and the SCTMA methods. 
As it can be seen, $A_{\down}(\bm{p},\omega)$ exhibits a sharp peak which is associated with the attractive polaron state and shows an almost quadratic behavior for increasing momentum of the impurities.
It is also apparent that the SCTMA spectrum (Figure~\ref{fig4}b) at low momenta is broadened when compared to the ETMA one (Figure~\ref{fig4}a) due to the induced beyond two-body interpolation correlations, e.g.,~three-body ones. 
At small impurity momenta, the~spectral peak of the attractive Fermi polaron within the present model as described by Equation~(\ref{eq:H}), is generally given by 
\begin{align}
A_{\down}(\bm{p},\omega)\simeq Z_{\rm a}\delta\left(\omega+\mu_\down-\frac{p^2}{2m_{\rm a}^{*}}-E_{\rm P}^{\rm (a)}\right),
\end{align}
where $Z_{\rm a}$ and $m_{\rm a}^*$ are the quasiparticle residue~\cite{Massignan2014} and the effective mass of the attractive polaron, respectively. 
At unitarity it holds that $Z_{\rm a}\simeq0.8$, $m_{\rm a}^*\simeq1.2$ m, and~$E_{\rm P}^{\rm (a)}\simeq-0.6E_{\rm F}$ within the zero-temperature and single-polaron limits~\cite{Combescot2007}. 
The behavior of these quantities has been intensively studied in current  experiments~\cite{Scazza2017,Nascimbene2009,Schirotzek2009} and an adequate agreement has been reported using various theories. 
For instance, Chevy's variational ansatz (being equivalent to the TMA at $T=0$ and $x\rightarrow 0$)~\cite{Chevy2006,Combescot2007} gives $Z_{\rm a}=0.78$, \mbox{$m_{\rm a}^*=1.17$ m}, and~$E_{\rm P}^{\rm (a)}=-0.6066E_{\rm F}$. More recently, the~functional renormalization group~\cite{Schmidt2011} predicts $Z_{\rm a}=0.796$ and $E_{\rm P}^{\rm (a)}=-0.57E_{\rm F}$, while according to the diagrammatic Monte Carlo method~\cite{Houcke2020} $E_{\rm P}^{\rm (a)}=-0.6157E_{\rm F}$. 
In this sense, nowadays, the~corresponding values of these quantities can be regarded as important benchmarks, especially for theoretical approaches.
{}{It is also worth mentioning that higher-order diagrammatic approximations such as the SCTMA do not necessarily lead to improved accuracy in terms of the values of relevant observables. 
In particular, a~detailed comparison between the predictions of the TMA and the SCTMA has been discussed in Ref.~\cite{Hu2018} demonstrating that the former adequately estimates the experimentally observed polaron energy whereas the SCTMA overestimates its magnitude in the strong-coupling regime. Moreover, the~diagrammatic Monte Carlo method based on bare Green's functions in self-energies exhibits a better convergence behavior compared to the ones employing dressed Green's functions due to the approximate cancellation of higher-order diagrams~\cite{Vlietinck2013}. 
As such, the~partial inclusion of higher-order diagrams by replacing the bare Green's functions with the dressed ones may lead to overestimating the molecule-molecule and the polaron-molecule scattering lengths in the strong-coupling regime~\cite{Tajima2019}.} 
\par

As we demonstrated previously (see Figure~\ref{fig3}), besides~the fact that the spectral response within the SCTMA method is broader compared to the one obtained in the ETMA, the~two spectra feature a qualitatively similar behavior. 
Indeed, both approaches evince that the spectra beyond $p=p_{\rm F}$ are strongly broadened. 
Recall that in this region of momenta the atoms of the majority component, which form the Fermi sphere, cannot follow the impurity atoms. 
This indicates that the dressed polaron state ceases to exist due to the phenomenon of the Cherenkov instability{~\cite{Grusdt2018,Nielsen2019}}, where the polaron moves faster than the speed of sound of the medium and consequently it becomes unstable against the spontaneous emission of elementary excitations of the medium. 
Such a spectral broadening can also be observed in mesoscopic spin transport measurements~\cite{Sekino2020} and may also be related to the underlying polaron-Boltzmann gas transition~\cite{Yan2019} since the contribution of high-momentum polarons can be captured in rf spectroscopy due to the thermal broadening of the Fermi distribution function in Equation~(\ref{eq:Ie}) at high temperatures.
Moreover, the~momentum-resolved photoemission spectra would reveal these effects across this transition. 
\par

{\begin{figure}[t]
\includegraphics[width=16cm]{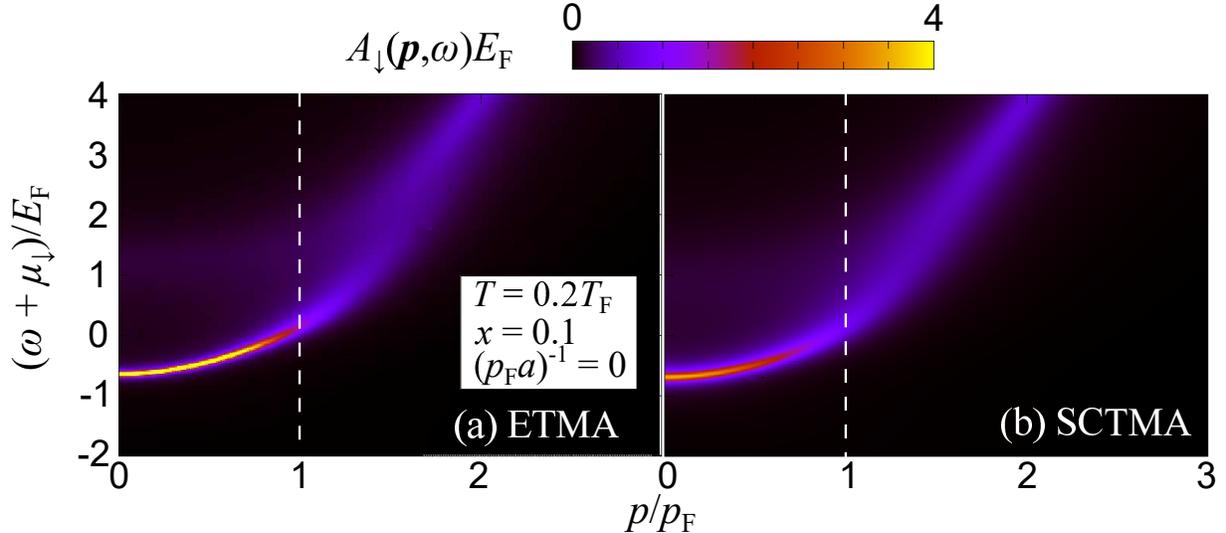}
\caption{Polaron
 spectral function $A_{\down}(\bm{p},\omega)$ as a function of the momentum $\bm{p}$ and the energy $\omega$ of the impurities at temperature $T=0.2T_{\rm F}$, impurity concentration $x=0.1$, and~interaction $(p_{\rm F}a)^{-1}=0$. 
$A_{\down}(\bm{p},\omega)$ is calculated within (\textbf{a}) the ETMA and (\textbf{b}) the SCTMA approaches. 
The vertical dashed line marks the Fermi momentum $p=p_{\rm F}$ of the medium.
While the two approaches predict qualitatively similar spectra with a sharp peak at low momenta and broadening above $p=p_{\rm F}$, the~SCTMA result (\textbf{b}) shows a relatively broadened peak at low momenta compared to the ETMA one (\textbf{a}).  
}
\label{fig4}
\end{figure}

We remark that the medium spectral function $A_{\up}(\bm{p},\omega)$ is also useful to reveal the properties of strong-coupling polarons in the case of finite temperature and impurity concentration.
Figure~\ref{fig_addfor3d} presents $A_{\up}(\bm{p},\omega)$ for various impurity-medium couplings ($(p_{\rm F}a)^{-1}=-0.4$, $0$, $0.4$, $0.7$, and~$1.0$) at $T=0.4T_{\rm F}$ and $x=0.1$.
At $(p_{\rm F}a)^{-1}=-0.4$ and $(p_{\rm F}a)^{-1}=0$,
$A_{\up}(\bm{p}=\bm{0},\omega)$ features a single peak at $\omega+\mu_\up=0$.
On the other hand, at~intermediate couplings $(p_{\rm F}a)^{-1}=0.4$ and $(p_{\rm F}a)^{-1}=0.7$, besides~a dominant spectral maximum a second peak appears around $\omega+\mu_{\up}=E_{\rm F}$. 
The latter evinces the backaction from the repulsive polaron because the inset of Figure~\ref{fig_addfor3d} shows that the repulsive polaron is located around $\omega+\mu_{\up}\simeq E_{\rm F}$.
Moreover, at~$(p_{\rm F}a)^{-1}=1$, another peak emerges in the low-energy region ($\omega+\mu_{\up}\simeq-3E_{\rm F}$).
This low-energy peak elucidates the emergence of two-body molecules with the binding energy given by $E_{\rm b}=1/(ma^2)$ due to the strong impurity-medium attraction.
Concluding, the~spectral function of the medium atoms can provide us with useful information for the recently observed smooth crossover from polarons to molecules~\cite{Ness2020}.
{}{Notice also that spectral and thermodynamic signatures of the polaron-molecule transition have been recently reported within a variational approach~\cite{Parish2021}, while the associated molecule-hole continuum can be captured using the TMA method~\cite{Schmidt2012}.}

\par
In the following, we shall elaborate on the behavior of the spectral function of lower dimensional Fermi polarons solely within the TMA approach. 
The latter provides an adequate description of the polaron formation in our case since the induced interpolaron interaction~\cite{MistakidisPPS,Mistakidisinduced} is weak in the considered mass-balanced~system. 

{\begin{figure}[t]
\includegraphics[width=11cm]{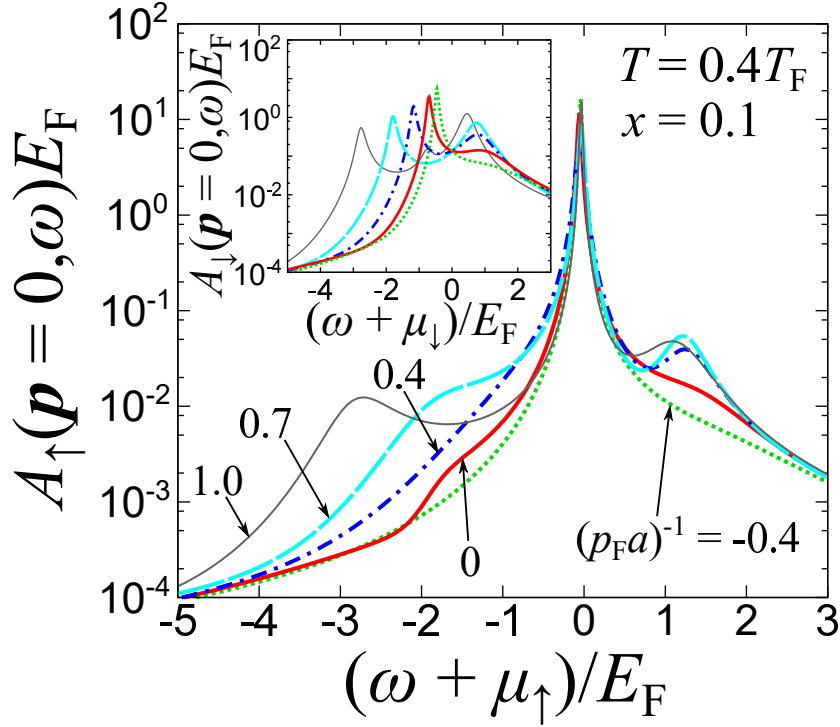}
\caption{Spectral
 function of the medium $A_{\up}(\bm{p}=\bm{0},\omega)$ within the ETMA approach at zero momentum of the impurity and for different impurity-medium couplings $(p_{\rm F}a)^{-1}=-0.4$, $0$, $0.4$, $0.7$, and~$1.0$.
The temperature and the impurity concentration are given by $T=0.4T_{\rm F}$ and $x=0.1$, respectively.
The inset shows the corresponding impurity spectral functions $A_{\down}(\bm{p}=\bm{0},\omega)$.
While the sharp peak at $\omega+\mu_\up\simeq0$ in $A_{\up}(\bm{p}=0,\omega)$ is associated with the bare state,
the small amplitude side peaks at positive ($\omega+\mu_\up\simeq E_{\rm F}$) and negative energies ($\omega+\mu_\down\simeq -3E_{\rm F}$ for the case with $(p_{\rm F}a)^{-1}=1$) originate from the backaction due to the impurities. } 
\label{fig_addfor3d}
\end{figure}

\subsubsection{Spectral Response of Fermi Polarons in~Two-Dimensions}

{In two spatial dimensions, the~attractive impurity-medium effective interaction \mbox{$g_{2D}<0$}} 
is always accompanied by the existence of a two-body bound state whose energy scales as $-1/(ma_{\rm 2D}^2)$~\cite{Klawunn2011}. 
Simultaneously, the~repulsive polaron branch appears at positive energies~\cite{Massignan2014} in addition to the attractive one located at negative energies. 
This phenomenology is similar to the case of a positive impurity-bath scattering length in 3D~\cite{Schmidt2012}. 
To elaborate on the typical spectrum of 2D Fermi polarons below we employ a homogeneous Fermi mixture characterized by an impurity concentration $x=0.1$, temperature $T=0.3T_{\rm F}$, and~a typically weak dimensionless coupling parameter $\ln(p_{\rm F}a_{\rm 2D})=0.4$ where $a_{\rm 2D}$ is the 2D scattering length introduced in Equation~(\ref{eq:a}).
The spectral response of both the fermionic background ($A_{\up}(\bm{p},\omega)$) and the impurities ($A_{\down}(\bm{p},\omega)$) for varying momenta and energies of the impurities within the TMA approach is depicted in Figure~\ref{fig5}. 
We observe that the small impurity concentration, i.e.,~$x=0.1$, leads to the non-interacting dispersion of the spectrum of the majority component given by $A_{\up}(\bm{p},\omega)\simeq\delta(\omega-\xi_{\bm{p},\up})$; see Figure~\ref{fig5}a.
In this case, therefore, the~medium does not experience any backaction from the impurities. 
Importantly, one can indeed identify a sizable backaction on the medium in the case of a larger impurity concentration and smaller impurity-medium 2D scattering length as shown in Figure~\ref{fig5}b1,b2 where $T=0.3T_{\rm F}$, $x=0.3$, and~$\ln(p_{\rm F}a_{\rm 2D})=0$. 
Moreover, since the repulsive interaction in the excited branch of the impurities ($\omega+\mu_\down\simeq E_{\rm F}$) 
is relatively strong, the~impurity excitation spectrum at positive energies $(\omega+\mu_\down>0)$ is largely broadened. 
We note that the stable repulsive polaron {branch} can be found in the case of small $a_{\rm 2D}$.  
It also becomes  evident that the impurity spectrum in 2D is largely broadened beyond $p=p_{\rm F}$ as compared to the 3D spectral response (Figure~\ref{fig4}).
Simultaneously, the~intensity of the metastable {impurity} excitation in the repulsive {branch} becomes relatively strong in both the 2D and 3D cases.
{This result implies that fast-moving impurities do not dress the medium atoms and occupy the {non-interacting} excited states in such high-momentum~regions.}

{\begin{figure}[t]
\includegraphics[width=13.5cm]{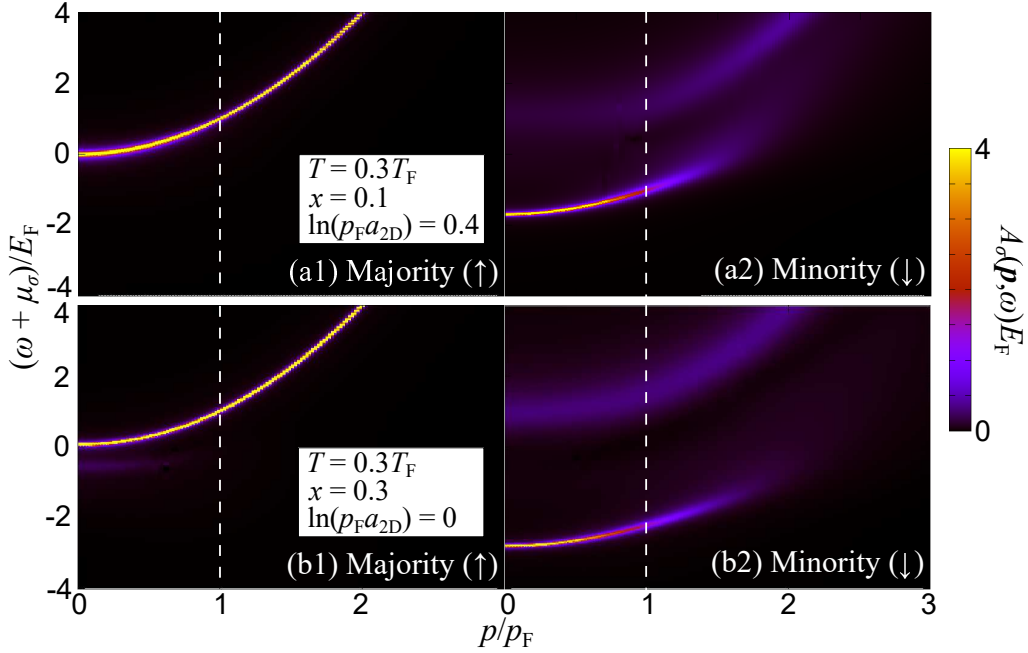}
\caption{Spectral
 function $A_{\sigma}(\bm{p},\omega)$ of the Fermi (\textbf{a1}) medium and (\textbf{a2}) impurities in two-dimensions for different momenta and energies of the impurities. 
We consider a temperature $T=0.3T_{\rm F}$, impurity concentration $x=0.1$, and~dimensionless coupling parameter $\ln(p_{\rm F}a_{\rm 2D})=0.4$.  
The vertical dashed line indicates the Fermi momentum $p=p_{\rm F}$ of the majority component atoms. 
While the majority component (a) exhibits a sharp peak with quadratic dispersion $\omega+\mu_\up=p^2/(2m)$, the~minority atoms (b) form the attractive polaron at negative energies ($\omega+\mu_\down<0$) and a broadened peak associated with the repulsive impurity branch at positive energies $(\omega+\mu_\down>0)$.
For comparison, we provide the spectral functions of the medium (\textbf{b1}) and the impurities (\textbf{b2}) in the case of $T=0.3T_{\rm F}$, $x=0.3$ and $\ln(p_{\rm F}a_{\rm 2D})=0$. 
Evidently, the~feedback on the medium from the impurities is enhanced in the low-momentum region ($p\simeq 0$). }  
\label{fig5}
\end{figure} 
\par
\subsubsection{Fermi Polarons in~One-Dimension}
   
In one spatial dimension the quasiparticle notion is somewhat more complicated as compared to the higher-dimensional case. 
Interestingly, various experiments are nowadays possible to realize 1D ensembles 
and thus probe the properties of the emergent quasiparticles. 
Below, we provide spectral evidences of 1D Fermi polarons and in particular calculate the respective $A_{\sigma}(\bm{p},\omega)$ (Figure~\ref{fig6}) for the background fermionic medium and the minority atoms within the $T$-matrix approach including the Hartree correction. 
The system has an impurity concentration $x=0.326$, it lies at temperature $T=0.157T_{\rm F}$, and~the 1D dimensionless coupling parameter for the impurity-medium attraction is $(p_{\rm F}a_{\rm 1D})^{-1}=0.28$ in Figure~\ref{fig6}(a1,a2).
For comparison, we also provide $A_{\sigma}(\bm{p},\omega)$ in Figure~\ref{fig6}(b1,b2) for the repulsive interaction case $(p_{\rm F}a_{\rm 1D})^{-1}=-0.55$ with system parameters $x=0.264$ and $T=0.598T_{\rm F}$.
We remark that the impurity-medium attraction is considered weak herein such that the induced interpolaron interactions are negligible.
In this sense, we do not expect significant deviations when considering the ETMA or even the SCTMA approaches.

{\begin{figure}[t]
\includegraphics[width=13cm]{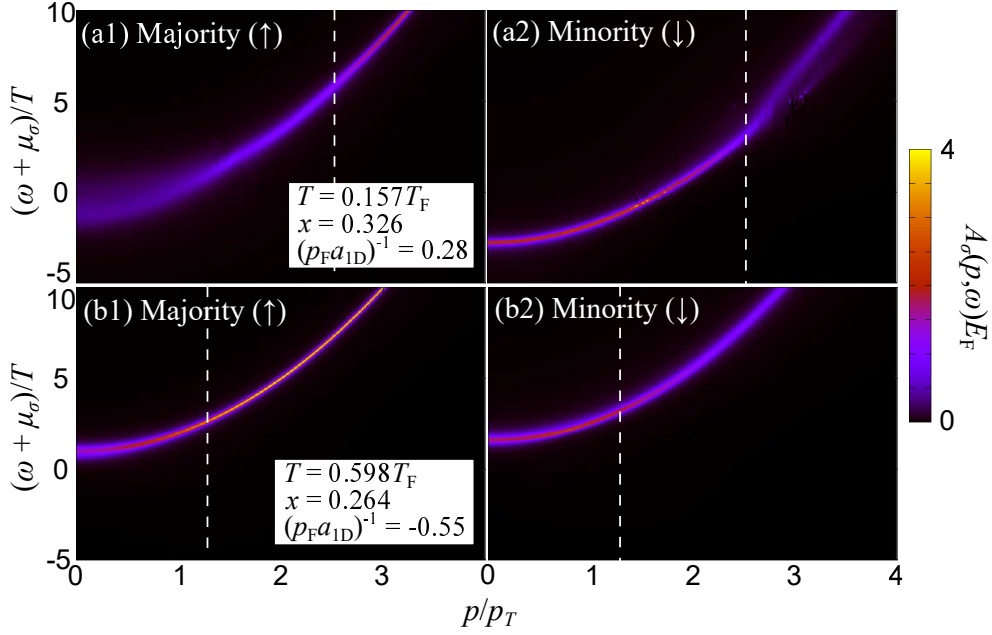}
\caption{Spectral
 function $A_{\sigma}(\bm{p},\omega)$ of the fermionic (\textbf{a1}) background and (\textbf{a2}) impurity atoms of concentration $x=0.326$ with an attractive medium-impurity interaction for varying momenta and energies of the impurities in one-dimension. 
The system is at temperature $T=0.157T_{\rm F}$ and dimensionless coupling parameter $(p_{\rm F}a_{\rm 1D})^{-1}=0.28$. 
$P_{T}=\sqrt{2mT}$ is the momentum scale associated with the temperature $T$.
The vertical dashed line marks the Fermi momentum $p=p_{\rm F}$ of the background atoms. 
The majority component (\textbf{a1}) is largely broadened due to the backaction from the impurities in the low-momentum region ($p\lesssim p_{T}$). 
On the other hand, the~minority component (\textbf{a2}) exhibits a sharp peak in the low-momentum region below $p=p_{\rm F}$ and it is broadened above $p=p_{\rm F}$.
For comparison, we show the (\textbf{b1}) medium and (\textbf{b2}) impurity spectral functions in the case of repulsive medium-impurity interaction characterized by $(p_{\rm F}a_{\rm 1D})^{-1}=-0.55$, where the temperature and the impurity concentraion are given by $T=0.598T_{\rm F}$ and $x=0.264$.
Although the impurity quasiparticle peak in the low-energy region ($\omega+\mu_\down\simeq 0$) is shifted upward, the~tendency of a spectral broadening is similar to the attractive case.   
}
\label{fig6}
\end{figure}
\par
It is also important to note here that in sharp contrast to higher spatial dimensions, the~coupling constant $g_{1D}$ does not vanish when  $\Lambda\rightarrow \infty$ in the renormalization procedure; see Section~\ref{sec2-1}. 
Thus, we take the Hartree shift $\Sigma_{\sigma}^{\rm H}=g_{1D}N_{-\sigma}$ into account in the building block of the self-energy diagrams~\cite{Tajima1D2020}.
This treatment is not necessary in the single-polaron limit since $\Sigma_{\up}^{\rm H}\rightarrow 0$ and $\Sigma_{\down}^{\rm H}\rightarrow g_{1D}T\sum_{\bm{p},i\omega_n}G_{\up}^0(\bm{p},i\omega_n)$ (which is included in the TMA self-energy) when $x\rightarrow 0$.
The non-vanishing coupling constant in 1D plays an important role in the emergence of induced interpolaron interactions as it has been recently demonstrated, e.g.,~in Refs.~\cite{Mistakidis2018,Kwasniok2020,Mukherjee2020}.
The polaronic excitation properties obtained within the TMA approach show an excellent agreement with the results of the thermodynamic Bethe ansatz~\cite{Guan2013}. 
The latter provides an exact solution in 1D and in the single-polaron limit at $T=0$~\cite{Klawunn2011,Doggen2013}. 
From these results, it is found that there is no transition but rather a crossover behavior between polarons and molecules.
As it can be seen by inspecting Figure~\ref{fig6}(a1) the spectrum of the majority component is affected by the scattering with the impurities. 
This is attributed to the relatively large impurity concentration $x$ considered here. 
In particular, $A_{\up}(\bm{p},\omega)$ is broadened at low momenta below $p=p_{\rm F}$.
On the other hand, the~spectral response of the impurities in \mbox{Figure~\ref{fig6}(a2)} exhibits a sharp peak associated with the attractive polaron below $p=p_{\rm F}$ and it becomes broadened above $p=p_{\rm F}$. Apparently, the~curvature of the position of the polaron peak corresponding to the effective mass (curvature of the dispersion) is changed around this value of the momentum.
Similar broadening effects of sharp peaks can be found even in the case of repulsive impurity-medium interaction shown in Figure~\ref{fig6}(b1,b2).
However, the~low-energy sharp peak (corresponding to the repulsive polaron) in the impurity spectrum (Figure~\ref{fig6}(b2)) is shifted to larger energies as a consequence of the impurity-medium~repulsion.


\section{Bose~Polarons}
\label{sec3}

In this section, we shall discuss 
the Bogoliubov theory of trapped Bose polaron systems in real space~\cite{Lampo2018, Mistakidis2019b, Takahashi2019}.
The reason for focusing on 
a real-space 
Bogoliubov theory is to elaborate on the deformation of the BEC medium in the presence of an impurity. 
Indeed, 
the interaction between the impurity and the medium bosons lead to significant inhomogeneities of the density distribution of the 
background 
which cannot be described within a simple Thomas--Fermi approximation. 
Such a modification of the boson distribution causes, for~instance, enhanced phonon emission
~\cite{MistakidisDiss,Mukherjee2020}.
Moreover, in~cold atom experiments the background bosons and the impurity are generally trapped.
Considering the impact of inhomogeneity that naturally arises in trapped systems, therefore, we treat the Bose polaron in real space without plane wave expansion because the momentum is not a good quantum number.
Below, we review the description of {a} Bose polaron in trapped systems at zero temperature using the
Bogoliubov theory and elaborate on the ground state properties. 
{}{We remark that our analysis, to~be presented below, is applicable independently of the shape of the external potential while for simplicity herein we consider the case of a harmonic trap.}

\par
In particular, we consider a 3D setting where a single atomic impurity is trapped 
in an external harmonic potential denoted by $V_\rI(r)$ and is embedded in a BEC medium that is also trapped in an another harmonic potential $V_\rB(r)$ whose center coincides with that of $V_\rI(r)$.
Hereafter, we use units in which $\hbar=1$. 
This system is described by the following model Hamiltonian 
{}{
\begin{equation}
\begin{split}
 \hat{H} 
 = &\intr \hat{\psi}^\dagger(\br) \ldk -\frac{\nabla^2}{2m_\rI}+V_\rI(\br)\rdk \hat{\psi}(\br)
 + g_\rIB \intr  \hat{\phi}^\dagger(\br) \hat{\phi}(\br) \hat{\psi}^\dagger(\br)\hat{\psi}(\br) \\
 &+ \intr \hat{\phi}^\dagger(\br) \ldk
    -\frac{\nabla^2}{2m_\rB}+V_\rB(\br) + g_{\rm BB}\hat{\phi}^\dagger(\br)\hat{\phi}(\br) \rdk \hat{\phi}(\br).
\end{split}
\end{equation}
}

Here, 
$\hat\phi$ and $\hat\psi$ are the field operators of the bosonic medium and the impurity, respectively.
$m_{\rI(\rB)}$ is the mass of 
the impurity atom (the medium bosons) and $\mu$ is the chemical potential of the 
medium bosons. 
The effective couplings $g_\rIB$ and $g_\rBB$ refer to the impurity-boson and boson-boson interaction strengths, respectively. 

\subsection{Bogoliubov Theory for Bose Polaron~Problems}
First, we calculate the expectation value of the Hamiltonian 
in terms of the single-impurity state 
$\ket{\rm imp}=\hat{a}_{\rm imp}^\dagger \ket{0}_{\rm imp}$ in order to integrate out the impurity's \mbox{degree-of-freedom}
{}{
\begin{equation}
\begin{split}
\hat{\mathcal{H}}_\rB
=& \intr \psi^*(\br)
    \ldk -\frac{\nabla^2}{2m_\rI} + V_\rI(\br)
    \rdk \psi(\br)
 \\&+
    \intr \hat{\phi}^\dagger(\br) \ldk -\frac{\nabla^2}{2m_\rB}+V_\rB(\br) 
+ g_\rIB |\psi(\br)|^2
+ g_{\rm BB}\hat{\phi}^\dagger(\br)\hat{\phi}(\br) \rdk \hat{\phi}(\br),
\label{Hb}
\end{split}
\end{equation}}
\noindent where $\hat{a}_{\rm imp}$ denotes the annihilation operator of {an impurity in} the ground state;  $\psi(r)$ is the corresponding wave function that {can be} determined self-consistently by \mbox{Equation~(\ref{eq:SD})}.
In this way, 
we have obtained the effective Hamiltonian for the medium bosons, in~which the bosons experience an effective potential constructed by the external trap and the density of the impurity $g_\rIB |\psi(\br)|^2$. 
Since we have set the temperature to zero in the present study, we have to assume {that the medium bosons possess} a condensed part, the~so-called order parameter or the macroscopic wavefunction, when using perturbation theory.
It is known~\cite{PethickSmith, PitaevskiiStringari, Dalfovo1999} that when BEC occurs, the~vacuum expectation value of the field operator $\hat\phi$ leads to a non-zero function which is used as an order parameter, i.e.,~$\langle \hat{\phi}(\br) \rangle_{\rm b}=\phi(\br)$, where $\langle \cdots \rangle_{\rm b}$ means $_{\rm b}\langle 0| \cdots |0\rangle_{\rm b}$. 
  The vacuum $|0\rangle_b$ is determined {from the effective Hamiltonian (\ref{Hb})} within the Bogoliubov theory to the second order of fluctuations. 
This is equivalent to splitting the operator as 
$\hat \phi = \phi + \hat \varphi$, where $\langle\hat \varphi\rangle_{\rm b} = 0 $.
Substituting this into the Hamiltonian of Equation~(\ref{Hb}) and expressing it in terms of the different orders of $\hat\varphi$, we can readily obtain the expansion  
$
 \hat{\mathcal{H}}_\rB
  \simeq {\mathcal H}^{(0)}+{\mathcal H}^{(1)}+{\mathcal H}^{(2)}
$
because the number of the non-condensed bosons is significantly smaller than that of the condensed ones at zero temperature and weak couplings.
In this expression, the~individual contributions correspond to
\begingroup
\makeatletter\def\f@size{9.5}\check@mathfonts
\def\maketag@@@#1{\hbox{\m@th\normalsize\normalfont#1}}%
\beq
{\mathcal H}^{(0)} &=&
\intr \psi^*
\ldk -\frac{\nabla^2}{2m_\rI} + V_\rI \rdk \psi
+
\intr \phi^*
 \ldk -\frac{\nabla^2}{2m_\rB} + V_\rB +g_\rIB |\psi|^2 +\frac{g_\rBB}{2} |\phi|^2 - \mu
\rdk \phi,
\\
{\mathcal H}^{(1)} &=&
\intr \hat{\varphi}^\dagger
\ldk
-\frac{\nabla^2}{2m_\rB} + V_\rB
+g_\rIB |\psi|^2
+g_\rBB |\phi|^2
-\mu
\rdk
\phi
+{\it h.c.},
\\
{\mathcal H}^{(2)} &=&
\frac12
\intr
\lk
\begin{array}{cc}
\hat{\varphi}^\dag & \hat{\varphi}
\end{array}
\rk
\lk
\begin{array}{cc}
\mathcal{L} & \mathcal{M} \\
\mathcal{M}^* & \mathcal{L}^*
\end{array}
\rk
\lk
\begin{array}{c}
\hat{\varphi} \\ \hat{\varphi}^\dag
\end{array}
\rk, \label{H2}
\eeq
\endgroup
where
$
	\mathcal{L}(\br)
	= -\frac{\nabla^2}{2m_\rB} + V_\rB(\br) +g_\rIB |\psi(\br)|^2 +2g_\rBB |\phi(\br)|^2 - \mu$, and~$\,
	\mathcal{M}(\br)=g_\rBB \phi^2(\br).
$
Note that we assume the weakly interacting limit of the medium to ensure the BEC {dominating} condition and thus $g_\rBB$ is adequately small such that the perturbation theory is valid.
In~the above expansion we ignore the contributions stemming from the third- and fourth-order terms {in} the field operator assuming that they are negligible
for the same~reason.

\par
Subsequently, let us derive the corresponding equations of motion that describe the Bose-polaron system. 
From the Heisenberg equation, the~bosonic field operator $\hat{\varphi}$ satisfies
$
 i\partial_t \langle\hat\varphi\rangle_{\rm b}
 = \langle[\hat\varphi, \hat{\mathcal H}^{(1)}+\hat{\mathcal H}^{(2)}] \rangle_{\rm b}=0
$
in the interaction picture.
Accordingly, it is possible to retrieve the celebrated Gross-Pitaevskii equation describing the BEC background 
\beq
\ldk
-\frac{\nabla^2}{2m_\rB} + V_\rB(\br)
+ g_\rIB |\psi(\br)|^2 + g_\rBB |\phi(\br)|^2 - \mu
\rdk
\phi(\br)=0.
\label{eq:GP}
\eeq

We remark that here, for~simplicity, we consider the stationary case where the condensate is time-independent.
Next, by~following the variational principle for $\psi$ namely $\delta\langle {\mathcal H}_\rB\rangle_{\rm b}/\delta \psi^*=0$, we arrive at the Schr{\"o}dinger equation for the impurity wavefunction 
\beq
\ldk
-\frac{\nabla^2}{2m_\rI} + V_\rI(\br)
 + g_\rIB|\phi(\br)|^2
 + g_\rIB n_{\rex}(\br)
\rdk \psi(\br) =0,
\label{eq:SD}
\eeq
where
$n_{\rex}(\br)=\langle\hat{\varphi}^\dagger(\br)\hat{\varphi}(\br) \rangle_{\rm b}$ is the density of the non-condensed bosons in vacuum, the~so-called {\it quantum depletion}.

To evaluate 
this expectation value, we need the ground state $\ket{0}_{\rm b}$ of the Hamiltonian that can be obtained by the diagonalization of Equation~(\ref{H2}), namely,
$
 H^{(2)} = \sum_n E_n \hat b_n^\dagger \hat b_n
$ is achieved using the following field expansion
$\hat{\varphi}(\br)=\sum_n\ldk \hat b_n u_n(\br)+\hat b_n^\dagger v_n^*(\br) \rdk$. 
Here the complete set $\{u_i, v_i\}$ 
satisfies the following system of linear equations being the so-called  Bogoliubov-de-Gennes (BdG) equations 
~\cite{Bogoliubov, deGennes}
\beq
\lk
\begin{array}{cc}
\mathcal{L}(\br) & \mathcal{M}(\br) \\
-\mathcal{M}^*(\br) & -\mathcal{L}(\br)
\end{array}
\rk
\lk
\begin{array}{c}
u_n(\br)\\
v_n(\br)
\end{array}
\rk
= E_n
\lk
\begin{array}{c}
u_n(\br)\\
v_n(\br)
\end{array}
\rk.
\label{eq:BdG}
\eeq

We remark that the BdG equations are commonly used in mode analysis of condensates. In~this context, the~real eigenvalues constitute the spectrum, {while} the complex eigenvalues unveil the dynamically unstable modes of the condensate~\cite{KatsimigaDAD,Katsimigaspinor}.
More precisely, if~complex eigenvalues exist then the Hamiltonian can not be expressed in the above-mentioned diagonal form in terms of the annihilation/creation operators. As~such, the~dynamically unstable situation is beyond the scope of the present description. 
By using this expansion, we can calculate the vacuum expectation,
e.g.,
$
 n_\rex(\br)=\sum_n |v_n(\br)|^2
$. 
For the numerical calculations, to~be presented below, the~total number of bosons 
$N_{\rm B}$ is conserved,~i.e.,~
\beq
 N_\rB = N_0 + N_\rex, \,\,\,\, {\rm with}\,\,\,\, 
 N_0 = \intr \, |\phi (\br)|^2 \,{\rm and}\, N_\rex=\intr \, n_\rex(\br). \label{constraint_number}
\eeq

This condition is achieved by tuning the chemical potential $\mu$ of the Bosonic medium. 
Notice that $N_ {\rm ex}$ becomes non-zero due to thermal fluctuations at finite temperature, while in the ultracold regime it can be finite due to the presence of quantum fluctuations, otherwise termed quantum depletion~\cite{Muellerfragm}.
We also remark that all of the above Equations~(\ref{eq:GP})--(\ref{eq:BdG}) need to be solved simultaneously.
The above-described treatment will be referred to in the following as the real-space formulation of the Bose-polaron problem.
\par

\subsection{Quantum Depletion {around a Bose~Polaron}}

Since $N_{B}$ is fixed (Equation~(\ref{constraint_number})), the~number of condensed particles $N_0$ changes due to the existence of $N_\rex$. 
This is a quantum effect that occurs even at zero temperature, and~it is called quantum depletion~\cite{Dalfovo1999}. 
We need to clarify that the term quantum depletion refers to the beyond mean-field corrections for the description of the bosonic ensemble.
In the following, we shall investigate the effect of {an}
impurity on the quantum depletion 
of the medium bosons at zero temperature. 
Indeed, the~quantum depletion is {a measurable} quantum effect that is included in 
Equation~(\ref{eq:SD})
and its quantification makes it possible to evaluate the 
backaction of the impurity {on} the medium condensate. 
\par
A commonly used external confinement in cold atom experiments is the harmonic potential.
As such, here, we consider that the traps of the impurity and the bosonic medium are spherically symmetric, namely,
\beq
	V_{\rB}(r)	= \frac{1}{2} m_\rB \omega_\rB^2 r^2 \,\,\,\,{\rm and}~~~~
	V_{\rI}(r)	= \frac{1}{2} m_{\rI} \omega_\rI^2 r^2.
\eeq

Accordingly, the~order parameter of the BEC and the impuritys' wave function have spherically symmetric forms, and~therefore the underlying BdG eigenfunctions are separable with the help of spherical harmonics {as}
\beq
    \phi(\br) = \phi(r), \,\,\,\,
    \psi(\br) = \psi(r), \,\,\,\,
    \ltk
    \begin{matrix}
        u_{n_r \ell m}(\br)\\
        v_{n_r \ell m}(\br)
    \end{matrix}
    \rtk=\ltk
    \begin{matrix}
        \mathcal{U}_{n_r\ell}(r)\\
        \mathcal{V}_{n_r\ell}(r)
    \end{matrix} \rtk
    Y_{\ell m} (\theta_1, \theta_2)  \,,
\eeq
where
 $r=|\br|$. 
Here, $(n_r, \, \ell, \, m)$ denote the
radial, azimuthal, and~magnetic quantum numbers, respectively.  

As a further simplification, we consider the situation where $\omega_\rI$ is sufficiently larger than $\omega_\rB$, namely, the~impurity is more tightly confined than the medium bosons. 
As such, the~order parameter $\phi$ of the condensate changes much more gradually with respect to the spatial change of the impurity's wave function $\psi$.
Since the {impurity's} wave function is relatively narrow compared to the condensate and the impurity-medium interaction is weak, the~impurity essentially experiences to a good approximation an almost flat (homogeneous) environment. 
This also means that trap effects are not very pronounced in this case. 
In this sense, $\phi$ can be regarded as being constant and the impurity's wave function can be well approximated by a Gaussian function i.e.,
$
 \psi(r)\simeq\lk \frac{\pi}{m_\rI \omega_\rI} \rk^{-\frac{3}{4}} \exp\lk -\frac{m_\rI\omega_\rI}{2} r^2 \rk
$.
{}{We remark that in the presence of another external potential, e.g.,~a double-well, one naturally needs to employ another appropriate initial wavefunction ansatz for the impurity.}
To experimentally realize such a setting it is possible to consider a $^{40}$K
Fermi impurity immersed in a $^{87}$Rb BEC, where $m_\rI/m_\rB \simeq 0.460$.  
For the medium we employ a total number of bosons $N_\rB = 10^5$ and the ratio of the strength of the trapping potentials $\omega_\rI/\omega_\rB = 10$ with
$\omega_\rB = 20 \times 2\pi \, {\rm Hz}$~\cite{Hu2016}.
Moreover, for~the boson-boson and impurity-boson interactions, we utilize the values  
$
 1/(a_\rBB n_\rB^{1/3}) = 100
$
and
$
 1/(a_\rIB n_\rB^{1/3}) = \pm 1
$
with
$
 n_{\rB}
  = N_{\rB} \Big/ \lk \frac{4\pi}{3} d_\rB^3 \rk
$ and
$
 d_\rB = \sqrt{m_{\rB}\omega_{\rB}}
$. 

To reveal the backaction of the impurity on the bosonic environment we provide the corresponding ground state density profiles of the condensed and the depleted part of the bath in Figure~\ref{fig:nex_r}a,c, respectively. 
In the case of $g_\rIB > 0$ ( $g_\rIB < 0$), the~condensate {experiences an additional potential hump (dip) at} the location of the impurity and eventually it {seems to be} slightly repelled from (pulled towards) the impurity as shown in Figure~\ref{fig:nex_r}b, 
where the deformation of the radial profile of the condensate from the case of zero impurity-medium interactions is provided. 
Moreover, in~order to appreciate the role of the quantum depletion of the BEC environment we illustrate its depletion density in the absence and in the presence of the impurity in Figure~\ref{fig:nex_r}b,d, respectively. 
Apparently, the~degree of the quantum depletion decreases (increases) (Figure~\ref{fig:nex_r}d) for $g_\rIB>0$ ($g_\rIB<0$), a~phenomenon that is accompanied by the deformation of the condensate density. 
The effect of the impurity on the quantum depletion of the condensate is summarized in the Table~\ref{table:delta_Nex}. 
Inspecting the latter we can deduce that the quantum depletion decreases (increases) when the interaction is repulsive (attractive).
This is a non-trivial result caused by the presence of the trap since in a uniform system{}{~\cite{Shchadilova2016,Drescher2020,Guenther2021}} the depletion always increases irrespectively of whether the interaction is positive or negative.

{\begin{figure}[t]
\includegraphics[width=15cm]{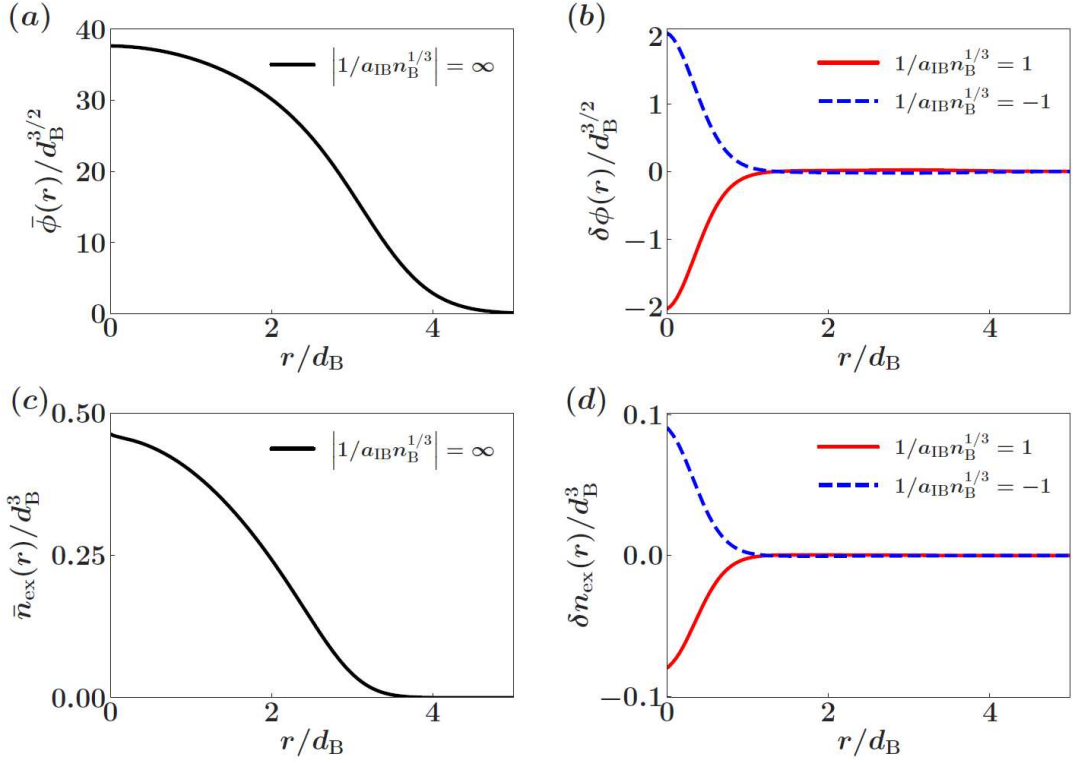}
\caption{Radial profiles of (\textbf{a}) the order parameter $\bar{\phi}(r)=\phi(r;g_\rIB=0)/\sqrt{N_0/4\pi}$ and (\textbf{c}) the density of depletion $\bar{n}_\rex(r)=n_\rex(r;g_\rIB=0)$ in the absence of an impurity. Differences of the radial profiles of (\textbf{b}) the order parameter $\delta \Phi(r)=(\phi(r;g_\rIB)-\phi(r;g_\rIB=0))/\sqrt{N_0/4\pi}$ and (\textbf{d}) the density of depletion $\delta n_\rex(r)=n_\rex(r;g_\rIB)-n_\rex(r;g_\rIB=0)$ in the presence of an impurity from the result depicted in (\textbf{a}) and (\textbf{c}), respectively.}
\label{fig:nex_r}
\end{figure}

\begin{table}[!ht]
\caption{The number of depletion $N_{\rex}$ and its deviation {$\delta N_\rex = 4\pi\int \!dr\, r^2 \delta n_\rex(r)$} from the case of zero impurity-medium interaction. It is evident that degree of depletion increases (decreases) for attractive (repulsive) interactions.
}
\label{table:delta_Nex}
	\begin{center}
		\begin{tabular}{c|ccc} \hline
			$1/(a_{\rIB}n_\rB^{1/3})$& $\infty$ & +1 & -1 \\ \hline
			$N_\rex$ & 24.244 & 24.220 & 24.270 \\
			$\delta N_\rex$ & 0 & -2.361$\times$ $10^{-2}$ & 2.584$\times$ $10^{-2}$ \\
	 \hline
\end{tabular}
	\end{center}
\end{table}


\section{Conclusions}
\label{sec4}
In this work, we have discussed the existence and behavior of Fermi and Bose polarons that can be realized in ultracold quantum gases focusing on their backaction {on} the background medium. 
We have explicated three different diagrammatic approaches applicable to Fermi polarons in the homogeneous case. 
These include the TMA, the~ETMA, and~the SCTMA frameworks, where {the} ETMA considers induced two-body interpolaron interactions and {the} SCTMA includes two- and three-body ones. 
Importantly, we have explicitly derived the mediated two- and three-body interpolaron correlation effects as captured within the different diagrammatic approaches.
Although these induced interactions are weak in the considered mass-balanced Fermi polaron {systems,}
our framework can be applied to various systems such as mass-imbalanced Fermi polaron systems. Using this strong-coupling approach, we analyze the spectral response of the Fermi polaron in one- two- and three- spatial dimensions at finite temperature. 
It has been shown that the spectral function of the minority component exhibits a sharp polaron dispersion in the low-momentum region but it is broadened for higher momenta. 
Moreover, we argue that the spectral response reflects the character of majority atoms forming a Fermi sphere while a strong interaction between the majority and the minority atoms induces a two-body bound state between a medium atom and an impurity particle. 
The presence of this two-body bound state becomes more important in lower~dimensions. 

Next, we present the mean-field treatment of trapped Bose polarons in three-dimensions and analyze the role of quantum depletion identified by the deformation of the background density within the framework of Bogoliubov theory of excitations. 
A systematic investigation of the latter enables us to deduce that the repulsive (attractive) impurity-medium interaction, giving rise to repulsive (attractive) Bose polarons, induces a decreasing (increasing) condensate depletion captured by the deformation of the density distribution of the host. 
This effect is a consequence of the presence of the external confinement since for a homogeneous background the quantum depletion increases independently of the sign of the impurity-medium interaction.
Therefore, this result is considered as a particular feature of the trapped system.

Our investigation opens up the possibility for further studies on various polaron aspects.  
In particular, the~effect of finite temperatures and the impurity concentration on the 2D Fermi polaron spectral response is expected to play a significant role close to the Berezinskii-Kosterlitz-Thouless transition of molecules~\cite{Tempere2009}. 
{}{Moreover, systems characterized by highly mass-imbalanced components, e.g.,~heavy polarons, provide promising candidates for the realization of more pronounced polaron-polaron induced interactions. 
However, the~treatment of these settings will most probably require a more sophisticated approach including for instance three-body correlations between the atoms of the medium. 
Additionally, the~investigation of finite sized systems at non-zero temperatures in the dimensional crossover from 3D to 2D as it has been reported e.g.,~in Ref.~\cite{Levinsen2012} but in the ultracold and single-polaron limits offers an interesting perspective for forthcoming endeavors.}
{}{Furthermore, the~comparison of the predictions of our methodology to treat the effect of quantum fluctuations in Bose polaron settings with other approaches based also on the mean-field framework~\cite{Drescher2020,Guenther2021} is certainly of interest.}
Finally, the~backaction of the impurities on the medium when considering dipolar interactions between the medium atoms may affect the density collapse of the medium at strong impurity-medium attractions~\cite{Nishimura2020} {}{and thus provide another intriguing prospect.}

\acknowledgments
The authors thank K. Nishimura, T. Hata, K. Ochi, T. M. Doi, and S. Tsutsui for useful discussion. S. I. M. gratefully acknowledges financial support in the framework of the Lenz-Ising Award of the University of Hamburg. 
This research was funded bu a Grant-in-Aid for JSPS fellows (Grant No. 17J03975) and for Scientific Research from JSPS (Grants No. 17K05445, No. 18K03501, No. 18H05406, No. 18H01211, and No. 19K14619).


\end{document}